\documentclass{article}
\usepackage{graphicx} 
\usepackage{wrapfig}
\usepackage{amsmath,amssymb}
\usepackage{subcaption}
\usepackage[margin=1in]{geometry}
\DeclareMathOperator{\E}{\mathbb{E}}
\usepackage{booktabs}
\usepackage{multirow}
\usepackage[flushleft]{threeparttable}
\usepackage{chngpage}
\usepackage[backend=biber,style=numeric,sorting=none]{biblatex}
\usepackage{makecell}
\usepackage{soul}
\usepackage{color}
\usepackage{float}
\usepackage{authblk}

\newcommand\blfootnote[1]{
  \begingroup
  \renewcommand\thefootnote{}\footnote{#1}
  \addtocounter{footnote}{-1}
 \endgroup
}

\title{Importance of hyper-parameter optimization during training of physics-informed deep learning networks}
\author[1,2]{Ashley Lenau}
\author[3,1]{Dennis M. Dimiduk}
\author[1,4]{Stephen R. Niezgoda}

\affil[1]{Department of Materials Science and Engineering, The Ohio State University, Columbus, Ohio, 43210}

\affil[2]{MST 8, Los Alamos National Laboratory, Los Alamos, New Mexico,87544}

\affil[3]{BlueQuartz Software, Springboro, Ohio, 45066}

\affil[4]{Department of Mechanical and Aerospace Engineering, The Ohio State University, Columbus, Ohio, 43210}
\begin{document}

\maketitle

\abstract{
Incorporating scientific knowledge into deep learning (DL) models for materials-based simulations can constrain the network's predictions to be within the boundaries of the material system. Altering loss functions or adding physics-based regularization (PBR) terms to reflect material properties informs a network about the physical constraints the simulation should obey. The training and tuning process of a DL network greatly affects the quality of the model, but how this process differs when using physics-based loss functions or regularization terms is not commonly discussed. In this manuscript, several PBR methods are implemented to enforce stress equilibrium on a network predicting the stress fields of a high elastic contrast composite. Models with PBR enforced the equilibrium constraint more accurately than a model without PBR, and the stress equilibrium converged more quickly.  More importantly it was observed that independently  fine-tuning each implementation resulted in more accurate models. More specifically each loss formulation and dataset required different learning rates and loss weights for the best performance. This result has important implications on assessing the relative effectiveness of different DL models and highlights important considerations when making a comparison between DL methods. }

\section{Introduction}
\blfootnote{Corresponding Author: Stephen R. Niezgoda, niezgoda.6@osu.edu}
Deep Learning (DL) simulations have been shown to significantly decrease prediction times compared to numerical simulations such as finite element methods \cite{Mozaffar2019,CROOM2022104191,Peivaste2022,TEICHERT2019}, potentially accelerating experimental processes and analyses. However, with DL models, this decrease in time is often at the expense of simulation accuracy. One source for the larger error is likely due to small datasets. DL is a data-driven method, meaning that large amounts of data are required to adequately train the network. As the dataset size decreases, the amount of missing information increases and the dataset may no longer be representative. The network is not able to interpolate these information gaps, which results in larger errors. In material science, acquiring large amounts of experimental data is time-consuming and costly, which can make data collection for DL a daunting task. Using numerically-based simulations provides a faster, reliable alternative to experiments, but are often still too computationally expensive to generate the large datasets needed to train DL networks. Moreover, simulated datasets also come with their own errors that the network will likely learn in addition to the network's error. 

Incorporating scientific knowledge into a DL network mitigates network error by including a more complete representation of the materials physics within the model. This can be achieved by altering the training data \cite{Sepasdar2021,HERRIOTT2020,Mishra2021}, network architecture \cite{Pokharel2021RNN,Mozaffar2019,frankel2019}, or loss function \cite{Zhang2020,Shukla2022,Pun2019,GOSWAMI2020,Sepasdar2021,Shah2019,singh2018,cang2017,Yang2018,Bhaduri2021} to reflect the physics of the system. The network is then encouraged to constrain the predictions to be within the physical boundary conditions imposed on the simulation. Physics-informed loss functions have become increasingly common and are similar in idea to the loss of a physics-informed neural network (PINN) \cite{PINNS1}. This forces a model to adhere to a physical state by either directly constraining it to be within the physical bounds of that property or encouraging the model to learn the desired physics to lower its overall loss.

The training and tuning process of a DL model is nontrivial, and how this process differs when the loss function is changed to include physics is rarely discussed. There is also little discussion on how physics-informed loss functions affect the time and data required to train a network to convergence. Additionally, few studies compare different loss functions that underwent separate fine-tuning. Tuning hyper-parameters like learning rate, number of epochs, loss weights, etc., have a great impact on the outcome of the network performance. Ref. \cite{li2020rethinking} shows that when fine-tuning a model used for transfer learning, different types of datasets required different effective learning rates. Ref. \cite{hoffer2021} compared different finite element method (FEM) surrogate models, including a multi-layer perceptron (MLP) and a network having a physics-based regularization (PBR) term in the loss function. The authors tuned the number of layers, epochs, neurons, batch size, and activation functions separately for each model and found that the MLP had better performance and concluded that the PBR network was difficult to train. However, the authors did not discuss hyper-parameters such as learning rate or loss function weight, which, as we show later, have a significant effect on the network performance. Ref. \cite{nabian2018} compared various regularizers (dropout, L1, L2, and PBR) on a neural network that modeled Burgers' equation, and compared them to a network without any regularization. The learning rate and weight of the regularizer were tuned separately for each model, showing that their PBR method had the best performance. Ref. \cite{bischof2021} compared different methods of balancing loss weights in a PINN network. Hyper-parameter optimization was performed for each PINN method separately, and the authors showed that their Relative Loss Balancing with Random Lookback method for updating the weights generally had better convergence. In fact, the multiple loss terms typically in PINN models are known to compete with each other during training and is an active area of research \cite{XIANG2022,Karniadakis2021,nosrati2023}. Every aspect of a neural network requires careful tuning to result in the best performance and this includes the loss function. Ref. \cite{krishnapriyan2021} showed that the loss function landscape can become more complex with PINN losses and concludes that the failure of a PINN model is likely due to poor optimization.

Since altering the loss function results in a change in the loss landscape, a fair comparison of data and convergence can be made only when a model with and without a physics-informed loss is fine-tuned to their respective loss functions.

Design strategies, such as topology Optimization \cite{deaton2014survey,wu2021topology,chi2021universal} or Integrated Computational Materials Engineering \cite{allison2006integrated,panchal2013key,wang2019integrated} rely on the repeated exercising of computational models as part of an optimization loop. For more computationally expensive simulations, like finite element or spectral methods, this becomes computationally intractable and potentially limits the searchable design space. Refs. \cite{Sepasdar2021,CHEN2023} used encoder-decoder-type networks to predict the crack location of composites, with Ref. \cite{CHEN2023} showing that their model was orders of magnitude faster than the fast Fourier transform (FFT) simulations used for their training data. Recently, Ref. \cite{xu2023cracknet} used CrackNet to model the crack growth of composites at several strain steps. Ref. \cite{YANG2019} used a convolutional neural network (CNN) to predict the strain of a 3-dimensional (3D) high-contrast elastic composite at each voxel given the local neighborhood. Refs. \cite{YANG2021,Yang2021_} used Pix2Pix networks to predict the stress and strain fields of randomly generated composite structures, which would allow faster composite design exploration. Several other studies \cite{YANG2021,Yang2021_,CROOM2022104191,Feng2021,Raj2021,Bhaduri2022,Gupta2023,Shokrollahi2023} show the success of using Pix2Pix, U-Net \cite{ronneberger2015unet}, or a similar encoder-decoder structure to perform composite to stress/strain field translations. Ref. \cite{CROOM2022104191} shows that a U-Net can predict the stress fields of porous microstructures much more quickly than FFT-based finite element solvers and that U-Net can outperform models like StressNet \cite{Nie_stressNet} with fewer training parameters. The skip-connections in the U-Net pass low-frequency information from input to output. This makes it a highly effective network for translating the stress fields of a composite since it doesn't have to learn shared spatial features between the input and output. 

For this study, we used a Pix2Pix \cite{isola2018} network to predict the normalized stress fields of a high elastic contrast two-phase composite. Few studies incorporate physical knowledge into their DL models predicting the mechanical behavior of a composite. Ref. \cite{yu2022} uses a Q-learning scheme to create composite arrangements that best prevent crack propagation by including fracture toughness in the reward process. Ref. \cite{Sepasdar2021} and Ref. \cite{CHEN2023} use ``attention'' mechanisms in their loss function and model architecture, respectively, that focus the models' efforts on important features, like cracks. Ref. \cite{Sepasdar2021} found that their attention loss led to more accurate crack predictions than the typical mean absolute error loss. 

In this paper, PBR losses are implemented to enforce the stress equilibrium of composite stress fields generated by a Pix2Pix network. Several PBR losses are studied to see the effect of a PBR term in a loss function in a general sense and to see how performance, training, and tuning change with different implementations of the same constraint. Learning rates and loss weights are fine-tuned to each function separately, and compared to a baseline network without PBR terms. Once tuned, the amount of data needed to train and the convergence of each implementation is discussed.

\section{Methods}
\subsection{Dataset Generation}
A phase field model \cite{chen1996continuum} of spinodal decomposition was used to generate synthetic microstructure images. The decomposition was modeled by the classical Cahn-Hilliard model \cite{chen1996continuum},
\begin{equation}
    \frac{\partial c}{\partial t} = D \nabla^2 (c^3 - c - \kappa  \nabla^2 c)
\end{equation}
\noindent where $c$ is a concentration parameter indicating domains at $\pm 1$, $D$ is a diffusivity with units $length^2/time$ (normalized to unity for the simulation), and $\kappa$ is a parameter that captures an effective surface energy such that $\sqrt \kappa$ gives the width of the transition regions between domains. The volume fraction, controlled by modulating $c$, of the two phases was varied from 30-70\% in increments of 1\%,  $\kappa$  was varied from 10-50 in increments of 10. Initial conditions were  Gaussian random noise with mean of $c$ a small variance to represent thermal fluctuations in composition. Each combination of $\kappa$ and volume fraction was assigned three different random initial conditions (without repetitions) to make up a total of 1025 microstructure images of $128\times 128$ pixels.

Localized mechanical response of the composite microstructures, in the form of spatially varying stress fields, was simulated via an elasto-viscoplastic fast Fourier transform (EVP-FFT) micromechanical model \cite{lebensohn2012elasto,patil2021comparison}. The microstructures were considered as high elastic contrast composite. The phases were modeled as elastically isotropic with an elastic stiffness ratio of 20 (Young's modulus of 40 MPa for the compliant phase and 800 MPa for the stiff phase). The material was elastically deformed to $\epsilon_{22}=0.001$ under plane-stress boundary conditions. The output of the simulation was the spatially varying stress fields $\sigma_{11}$, $\sigma_{22}$, and $\sigma_{12}$. As a pre-processing step for machine learning, each set of stress fields were individually normalized using min-max normalization to be in a range of -1 to 1 for a given set. 

\subsection{Deep Learning Background}
\begin{figure}
    \centering
    \captionsetup{width=.9\linewidth}
    \includegraphics[width=\linewidth]{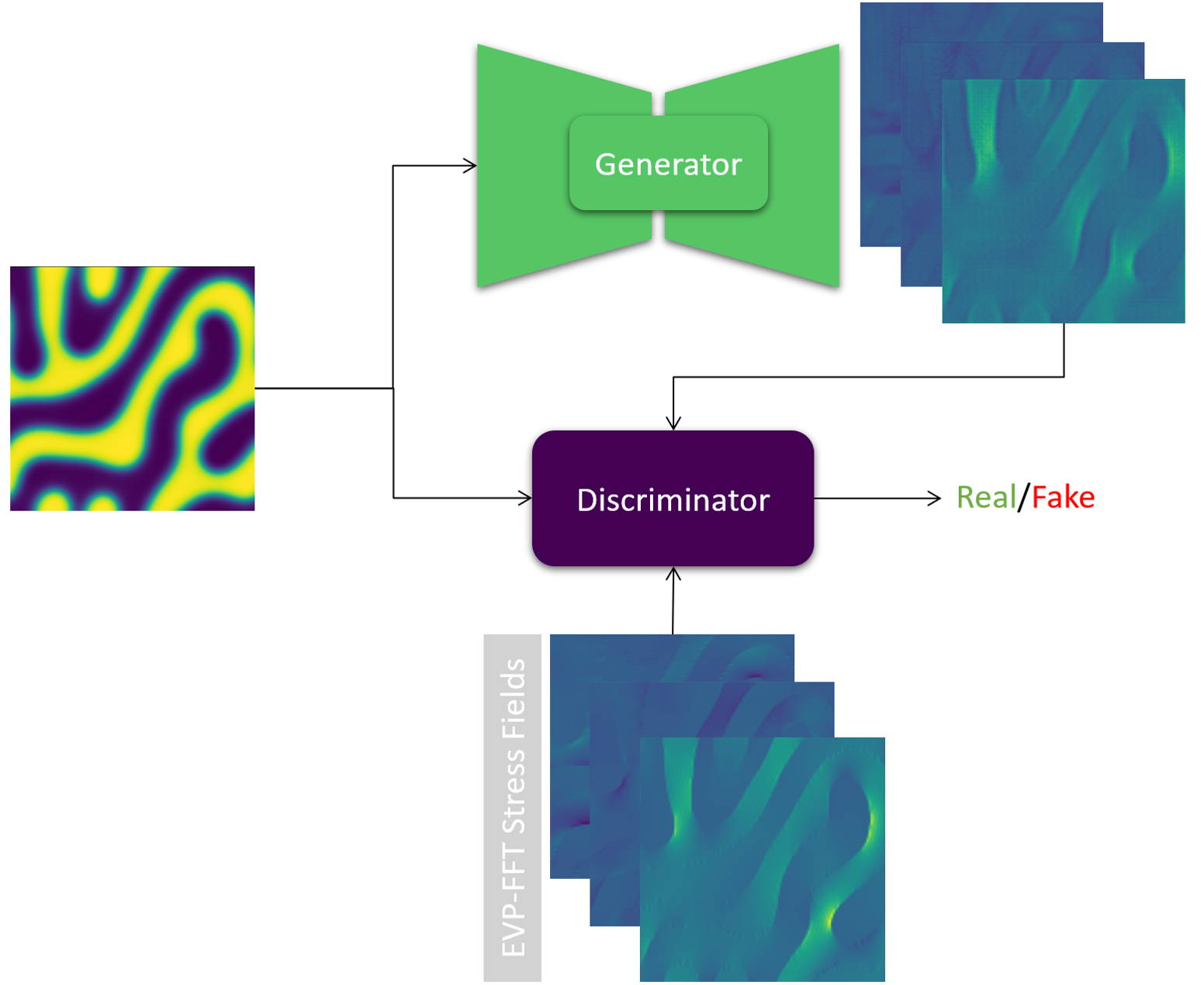}
    \caption{Schematic of a Pix2Pix GAN. The generator translates an image of a high elastic contrast composite to the corresponding stress fields. The discriminator determines the probability of a translation being real or fake based on network predictions and training data.}
    \label{Fig:1}
\end{figure}
A deep learning model was trained to predict the localized stress fields from the microstructure images. Within the broader context of machine learning this falls under the class of image translation \cite{I2I_rev}. A variant of the Pix2Pix network was utilized for this study due to  demonstrated success in other related image translation tasks \cite{isola2018}. Pix2Pix is a variation of a Generative Adversarial Network (GAN) model \cite{goodfellow2014} where the generator network has a U-Net architecture \cite{ronneberger2015unet}. U-Nets are a `staple in translation'-type networks because of their ability to pass on information from the input to the output through concatenations of corresponding encoding and decoding layers. This structure is highly efficient for image translation as the network does not need to learn redundant information that is shared between the input and output images. The discriminator network is a convolutional neural network that classifies an input as ``real'', meaning an actual training image, or ``fake'', meaning synthesized by the generator, on a patch basis. The objective function for the Pix2Pix network is the conditional GAN (CGAN) \cite{mirza2014} objective and is shown in Equation \ref{eq1},

\begin{multline}\label{eq1}
    V_{Pix2Pix}=\underbrace{min}_{G}\underbrace{max}_{D} \E_{X\sim P_{data}\ } [logD(X,Y)]
    + \E_{Z\sim P_{Z}\ } [log(1-D(G(Z,Y),Y))] \\
    + \beta L1(G(Z,Y))
\end{multline}

\noindent where $D$ and $G$ represent the discriminator and generator networks, respectively. $X$ represents an instance coming from the training data distribution (a set of stress fields), $P_{data}$, and $Y$ is the label (the two-phase composite image). $Z$ is a sampling from the noise distribution, $P_{Z}$, where $G(Z,Y)$ is the output of the generator model. For the Pix2Pix network, noise comes from the dropout layers in $G$ \cite{isola2018}. $\E$ represents the expected value. The $L1$ term is not part of the CGAN objective but is an L1-regularization term to encourage less blurring in predictions and $\beta$ is a weighting term. The workflow of the network is shown in Figure \ref{Fig:1}. The two-phase composite image is the input into the generator network and acts as the label. The stress fields in the 11-, 12-, and 22-directions are the outputs of the generator. The discriminator trains from both real and generated composite-to-stress field translations as input and assigns the translation a probability of being real \cite{isola2018}. Real stress field translations are assigned a ``1'' probability while generated predictions are assigned a ``0''. The discriminator is trained to predict these labels on a patch basis with a binary cross-entropy loss as shown in Equation \ref{eq1}. For our network, a patch size of 8x8 pixels was used. The generator learns from the discriminator's predictions and tries to trick the discriminator into assigning generated translations a higher probability. This objective will be used to compare to the PBR methods and will be referred to as the ``baseline'' method.

\subsection{Physics-Based Regularization Methods }
Three different loss functions were constructed to enforce stress equilibrium, each of which evaluates the divergence field computed from a set of stress fields.  Ignoring external body forces, this can be expressed mathematically by requiring the divergence of the stress field to be zero at each point in the material, $\nabla\cdot\sigma=0$. Given the plane stress boundary conditions the stress divergence can be expressed as: 
\begin{equation} \label{eq2}
K_1(\sigma)=\frac{\partial\sigma_{11}}{\partial x_1}+\frac{\partial\sigma_{12}}{\partial x_2},\ \ K_2(\sigma)=\frac{\partial\sigma_{12}}{\partial x_1}+\frac{\partial\sigma_{22}}{\partial x_2}
\end{equation}
\begin{figure}
    \centering
    \captionsetup{width=.9\linewidth}
    \includegraphics[width=0.3\linewidth]{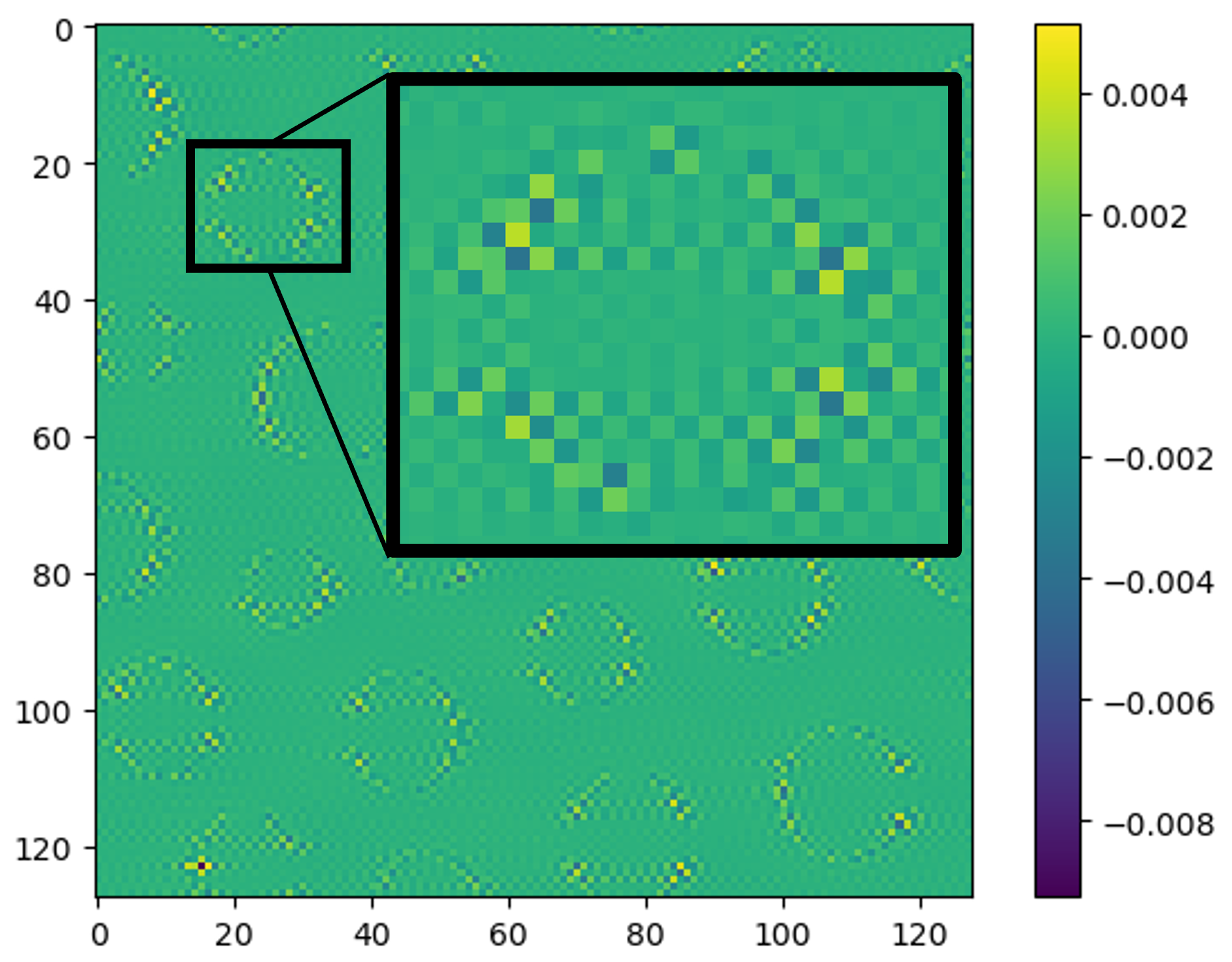}
    \caption{Divergence field from the training dataset. The zoomed-in portion shows greater deviations from equilibrium around a phase interface.}
    \label{fig:2}
\end{figure}
\noindent where, $K_{1}$ and $K_{2}$ is the divergence in the $x_{1}$- and $x_{2}$-direction, respectively, and $\sigma_{ij}$ is the Cauchy stress tensor. The equations listed in Equation \ref{eq2} are the foundation for the PBR terms discussed in the following sections. Several other publications have used this divergence condition to enforce stress equilibrium in their loss functions for similar applications \cite{bolandi2022,YONEKURA2023}.
 
The EVP-FFT simulation used to generate the dataset converged to a stress solution having a mean point-wise divergence value of $\sim 10^{-15}$ and a root-mean-square (RMS) value of $\sim10^{-3}$. RMS is particularly sensitive to outliers and is a strong metric to represent the equilibrium errors that are typically larger at the boundary interface between the two phases (see Figure \ref{fig:2}). These boundary errors are combinations of numerical residuals of solving for stress equilibrium under the constraints of the constitutive model, and artifacts of the FFT solver in the form Gibbs oscillations \cite{GibbsPhen} caused by the abrupt change in properties (and resulting stress fields) at the interface between the stiff and compliant phases in the microstructure. 

Three different methods to incorporate stress equilibrium were utilized in this study and will be described below. The regularization methods discussed here will be implemented on $D$ or $G$, and will either encourage ``true'' equilibrium (as with the simple addition method) or similar average equilibrium errors between the targets and predictions. The approaches taken for this paper are not exhaustive, and other physics-based regularization strategies can be imagined, such as calculating the pixel-wise error between the divergence fields from the EVP-FFT simulation and network predictions. However, we feel the methods selected for this study represent a useful cross-section of training and convergence behaviors and results.

\subsubsection{Simple Addition}

The most straightforward regularization method is the weighted addition of the absolute divergence values directly to the objective function, analogously to the L1 regularization implemented in the base Pix2Pix network as shown in Equation \ref{eq7}.
\begin{equation}\label{eq7}   
    \begin{split}
    V_{Add Div}= V_{Pix2Pix} + \gamma \left(|K_{1}(G(Z,Y))| + |K_{2}(G(Z,Y))|\right)
    \end{split}
\end{equation}

\noindent The divergence fields are calculated using Equation \ref{eq2} from the stress fields generated by the Generator network, $G$. As the divergence fields approach zero, little to no penalty is fed back to $G$. If the deviation is large, the model will be penalized by the magnitude of the divergence scaled by $\gamma$. In this way, the Simple Addition regularizer biases the Generator to synthesizing stress fields that are closer to equilibrium.

\subsubsection{Sigmoid Function}
This regularization method is implemented on the Discriminator network, $D$, and predicts the probability of whether a divergence field is calculated from a set of ``real'' or synthetic stress fields. More specifically, the RMS of a divergence field is evaluated to encourage predictions to have similar error convergence to the EVP-FFT simulation. Only the order of magnitude of the RMS value is considered to make the regularization method less constrictive (Equation \ref{eq3}). The sigmoid function in Equation \ref{eq4} describes the probability of an order of magnitude from Equation \ref{eq3} coming from the ground truth dataset.
\begin{equation}\label{eq3}
    M_{d} = log_{10}\sqrt{\frac{1}{N}\sum_{n}\mathrm({K}_{n}^{d}(\sigma))^2}
\end{equation}
\begin{equation}\label{eq4}
    S_{d} = -2 \left(\frac{1}{1+e^{-M_{d}}}-0.5\right)
\end{equation}
\noindent $M_{d}$ represents the RMS order of magnitude of a divergence field, with $d=1,2$ indicating the field from Equation \ref{eq2}, $N$ is the number of points in a divergence field, and $n$ is a point in the divergence field. If $M_{d}$ is close to -3, then the probability, $S_{d}$, of that divergence field being calculated from real stress fields is close to 1. The closer $M_{d}$ is to zero, the less likely it is to come from real stress fields and $S_{d}$ will be assigned a lower probability. The two probability values ($S_{1}$ and $S_{2}$ from $K_{1}$ and $K_{2}$, respectively) are multiplied by the original discriminator output to get the final discriminator output,
\begin{equation}\label{eq5}
D_{sig} = D(\sigma,Y)S_{1}S_{2}
\end{equation}
where $\sigma$ is a set of stress fields either from $G$ or the training dataset. The new objective function for the network becomes
\begin{multline}\label{eq6}
    V_{Sigmoid}=\underbrace{min}_{G}\underbrace{max}_{D} \E_{X\sim P_{data}\ } [logD_{sig}(X,Y)] 
    +\E_{Z\sim P_{Z}\ } [log(1-D_{sig}(G(Z,Y),Y))] \\
    + \beta L1(G(Z,Y))
\end{multline}
This new objective will encourage $G$ to produce plausible stress fields through the original Pix2Pix objective, and the additional probabilities on the $D$ network will encourage the $G$ network to produce stress fields that have plausible divergence fields as well.

\subsubsection{$\boldsymbol{Tan^{-1}}$}

This method is similar in spirit to the sigmoid method but is implemented on the generator, $G$. The RMS values from a generated divergence field need to be the same as the RMS from the target divergence field to have no additional penalty. A $tan^{-1}$ is applied for more stable gradient updates at larger errors. At smaller errors, the function becomes the absolute difference between the RMS values. The objective of this method becomes

\begin{equation}\label{eq8}
RMS_{d} = \sqrt{\frac{1}{N}\sum_{n}K_{n}^{d}(\sigma)^{2}}
\end{equation}

\begin{multline}\label{eq9}   
    V_{tan^{-1}}= V_{Pix2Pix} + \gamma (|tan^{-1}(RMS_{1}(G(Z,Y)))-RMS_{1}(X)| \\
    + |tan^{-1}(RMS_{2}(G(Z,Y)))-RMS_{2}(X)|)
\end{multline}
\noindent where $RMS_{d}(G(Z,Y))$ calculates the divergence RMS from generated stress fields and $RMS_{d}(X)$ from target stress fields. This method encourages the $G$ network to produce stress fields that have similar divergence errors to the training dataset.

\subsection{Network Specifics}

During the training process of a DL network, the trainable weights of a network are iteratively updated to minimize the network's loss function. The weights are updated by backpropagation \cite{backprop} of the error, where the gradient of the loss for each training weight is calculated. This gradient indicates the direction the weight needs to be updated to minimize the loss. The learning rate, $\alpha$, is a hyper-parameter defining how large of a step to take along that gradient for the update ($\alpha$ is sometimes referred to as step size \cite{kingma2017adam}). Making it too small may cause the model to get stuck in local minima, and conversely with too large an $\alpha$ the network may miss the global minimum. Changing or adding to the loss function changes the landscape of the loss \cite{krishnapriyan2021,basir2022critical}, resulting in new minima, and will likely require a learning rate adjustment. In the case of combined or regularized loss functions (like PBR), the loss weights need to be balanced so that the network doesn't focus too much, or too little on one of the terms, causing one metric to suffer at the expense of the other. For these reasons, tuning learning rates and loss weights was a focus area of this study.  

The network was tuned incrementally so that only one portion of the GAN was being optimized at a time. Throughout the tuning process, new parameters are evaluated by using the original Pix2Pix parameters as a benchmark. Adam optimization \cite{kingma2017adam} was used for all networks, which uses an adaptive effective learning rate for parameter updates and is the standard optimization procedure due to its efficiency. 

We found that the learning rate hyper-parameter, $\alpha$,  strongly effected the relative convergence of the different regularization methods. The learning rate for each regularization method was optimized by gradually increasing the learning rate throughout training within the range [0.00001,0.001] to maximize the rate of decrease in the loss \cite{smith2017cyclical} (the $\beta_1$ and $\beta_2$ parameters from Ref. \cite{kingma2017adam} were kept constant). This process is performed separately for the $G$ and $D$ networks where the default learning rate from Pix2Pix was retained for one network while the other network's learning rate was being tuned. The original Pix2Pix objective uses L1 regularization, but for this study L2 regularization was also tested for each model. The weight of the L1/L2 regularization and the PBR terms were tuned separately. For the L1/L2 and PBR losses on the generator, loss weights [10,50,100,200,1000] were initially tested, then iteratively updated until stress and equilibrium errors were both minimized, or a balance (pareto optimality) was found. The weight of the discriminator loss ($D-weight$) was also tuned within the range [0.5,1] in increments of 0.1.  

The tuning process was performed twice; once with $\sim 50\%$ of the training data and again with $\sim 75\%$ of the training data. This was to study the effect of tuning a network with smaller datasets. The validation dataset is used throughout the tuning process to verify and test hyper-parameters, and for final model selection. The test dataset isn't used until hyper-parameter optimization is completed and after the networks are completely trained to avoid any data leakage \cite{MAT_ML_best_prac}. 

After hyper-parameter optimization, each of the tuned models is trained with the full-sized training dataset. Each model was trained and independently re-trained in 10 separate training sessions. Training multiple instances of the networks over the same data was performed to give a better understanding of the expected performance if another user were to train the network on their own. Additional experiments  to study the effect of dataset size were performed by training with 25\%, 50\%, and 75\% of the training dataset (also averaged across 10 training sessions). For all training sessions, the network state was saved every 1000 iterations to record the performance throughout training and convergence. Unless stated otherwise, all results shown were generated using the test dataset from the saved checkpoints. 

The network was built using TensorFlow 2.26 \cite{tensorflow2015}. The Ohio Supercomputer was used to complete all runs using the Owens cluster having a 160 NVIDIA Tesla P100 \cite{OhioSupercomputerCenter1987}. The code was adapted from \cite{pix2pixCode} and code to reproduce all examples in this manuscript can be found at https://github.com/mesoOSU/PB-GAN.

\section{Results}
\subsection{Tuning}
\begin{figure}[H]
\centering
\captionsetup{width=0.9\linewidth}
\begin{subfigure}{\textwidth}
    \centering
        \includegraphics[width=\linewidth]{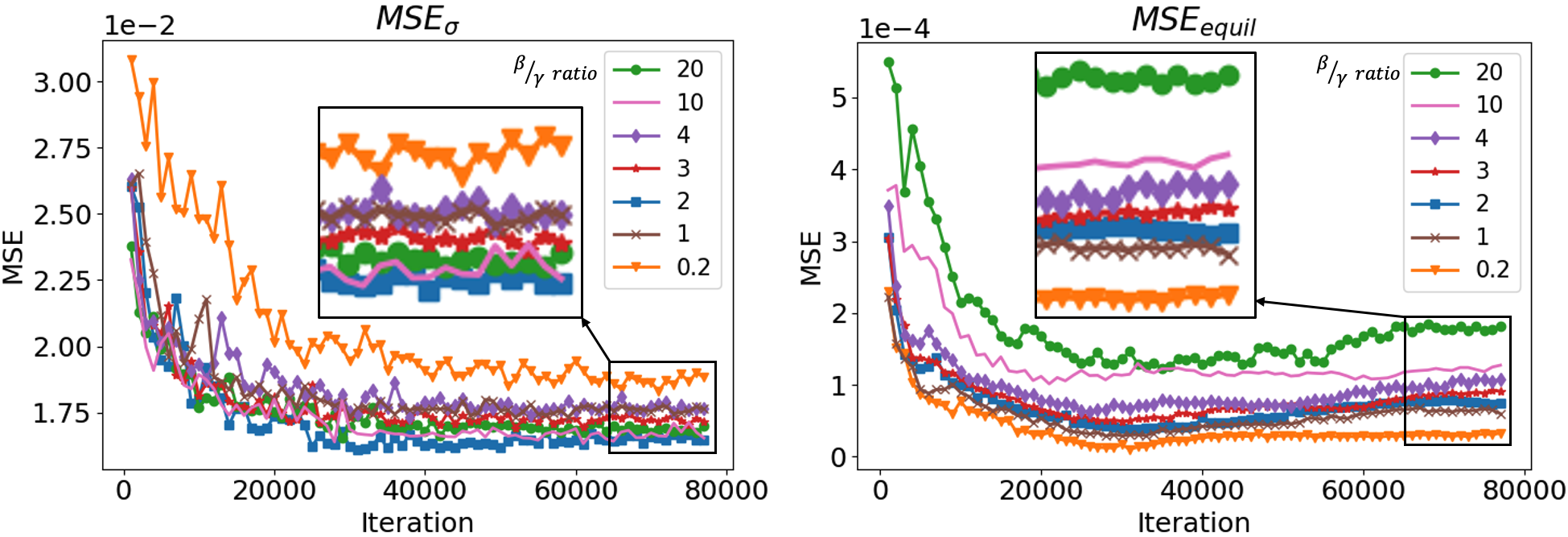}
        \caption{Varying $\beta$}
        \label{vary_L}
\end{subfigure}\vspace{0.5cm}

\begin{subfigure}{\textwidth}
    \centering
        \includegraphics[width=\linewidth]{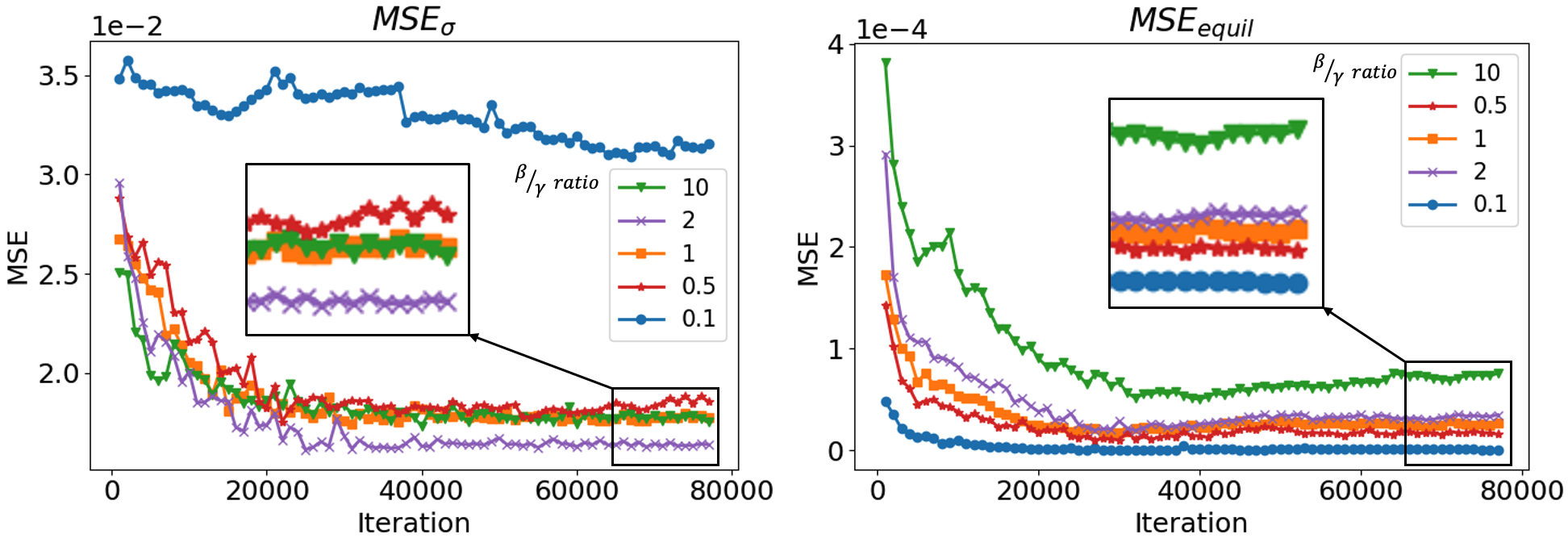}
        \caption{Varying $\gamma$}
        \label{vary_PB}
\end{subfigure}
\caption{$\text{MSE}_{\sigma}$ (left) and $\text{MSE}_{equil}$ (right) during training the simple addition model. (a) varies the weight of $\beta$ (L1 or L2 weight) where $\gamma=50$ and other weights and parameters are kept constant. (b) varies $\gamma$ (weight of the PBR) where $\beta=100$ and other weights and parameters are kept constant. The legend in each plot corresponds to the $\beta/\gamma$ ratio. Plots were generated using the validation dataset.}
\label{fig:L2_vs_Div}
\end{figure}

During hyper-parameter optimization, we found that the L1/L2 regularization and PBR terms often competed with each other during training. A similar loss competition has been observed and discussed by other authors \cite{scheinker2023,choi2022,Karniadakis2021}. The weights of these two terms need to be balanced so that neither the stress field (enforced by L1/L2) nor equilibrium (enforced by Equations \ref{eq6}, \ref{eq7}, or \ref{eq9}) errors suffered too greatly for the benefit of the other. In Figure \ref{fig:L2_vs_Div}, the mean squared error (MSE) of the stress fields ($\text{MSE}_{\sigma}$) and MSE of equilibrium ($\text{MSE}_{equil}$) is plotted throughout training for the simple addition method for various $\beta/\gamma$ ratios. Figure \ref{fig:L2_vs_Div}a varies $\beta$ and Figure \ref{fig:L2_vs_Div}b varies $\gamma$.
As the $\beta/\gamma$ ratio increases, $\text{MSE}_{\sigma}$ decreases generally at the expense of $\text{MSE}_{equil}$. 
When $\beta/\gamma = 20,10$, $\text{MSE}_{\sigma}$ is among the lowest errors, but the highest for $\text{MSE}_{equil}$ in Figure \ref{fig:L2_vs_Div}a. 
A similar trend is observed when increasing $\gamma$. When $\beta/\gamma=0.1$, $\text{MSE}_{\sigma}$ is the highest error, but has the lowest error for $\text{MSE}_{equil}$. A balance was found by choosing $\beta/\gamma=2$ where $\text{MSE}_{\sigma}$ is the lowest and in the middle for $\text{MSE}_{equil}$. Similar trends were observed for the other PBR terms, but the simple addition method displays them most clearly.

\begin{figure}[H]
\centering
\captionsetup{width=\linewidth}
\begin{subfigure}{0.4\textwidth}
    \centering
        \hspace*{-1.5in}
        \includegraphics[width=\linewidth]{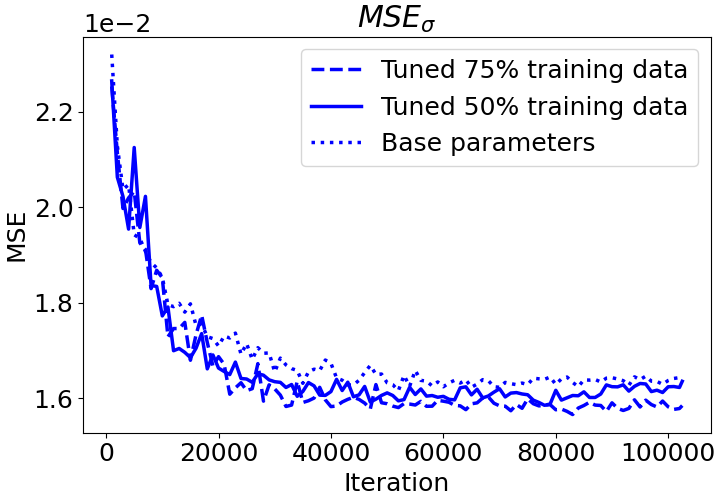}%
        \includegraphics[width=\linewidth]{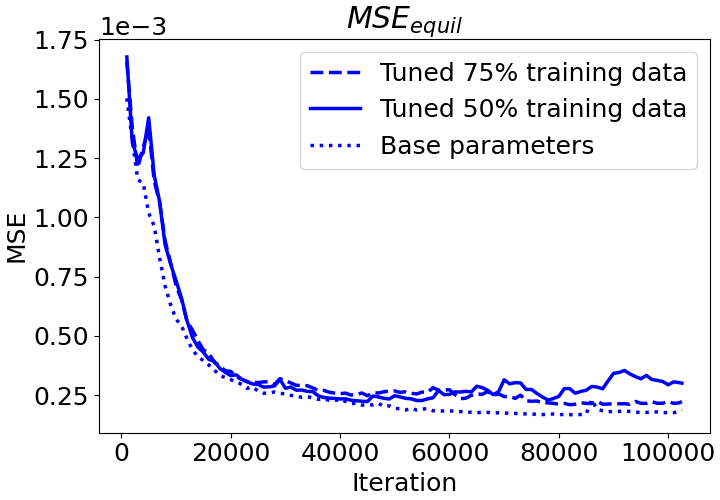}
        \caption{Baseline}
        \label{graph_tuning_NR}
\end{subfigure}

\begin{subfigure}{0.4\textwidth}
    \centering
        \hspace*{-1.5in}
        \includegraphics[width=\linewidth]{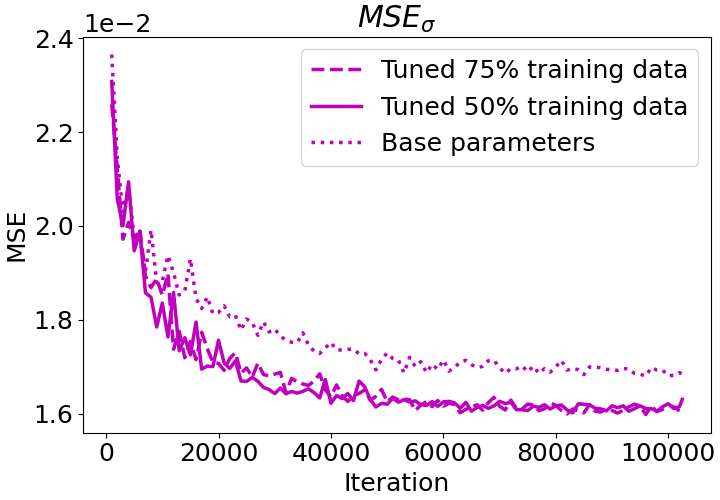}%
        \includegraphics[width=\linewidth]{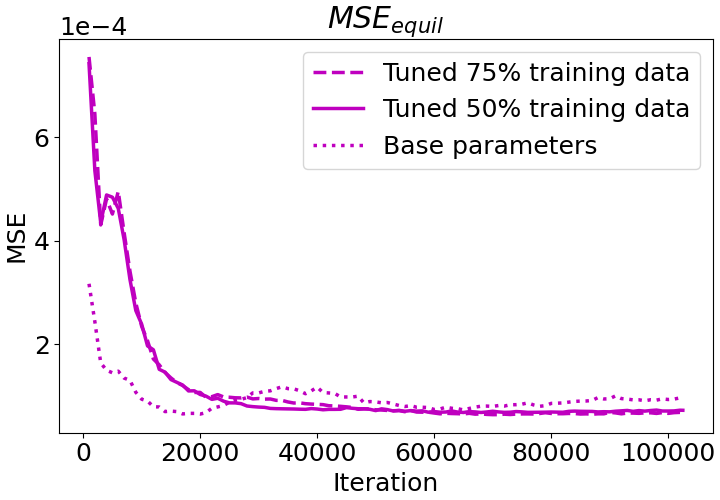}
        \caption{Sigmoid}
        \label{graph_tuning_Sig}
\end{subfigure}

\begin{subfigure}{0.4\textwidth}
    \centering
        \hspace*{-1.5in}
        \includegraphics[width=\linewidth]{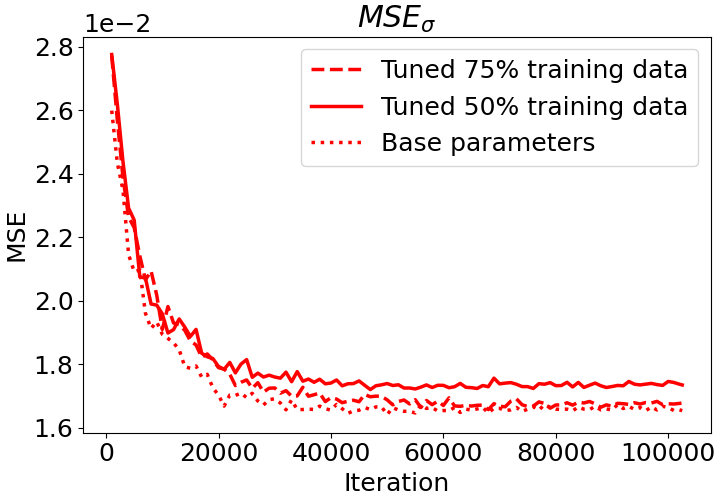}%
        \includegraphics[width=\linewidth]{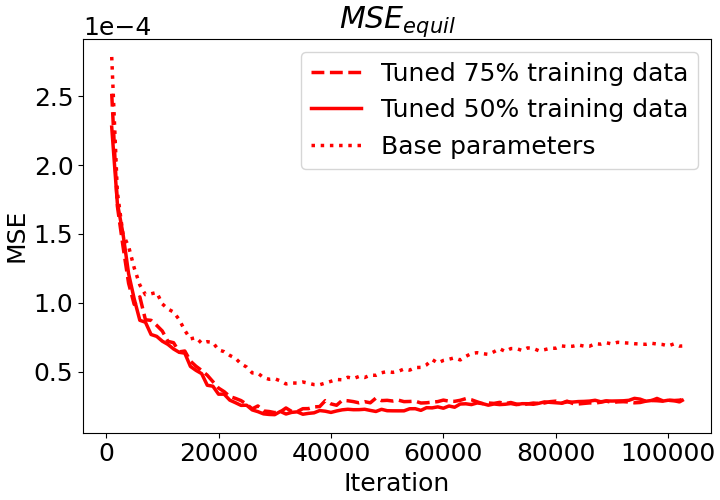}
        \caption{Simple Addition}
        \label{graph_tuning_AD}
\end{subfigure}

\begin{subfigure}{0.4\textwidth}
    \centering
        \hspace*{-1.5in}
        \includegraphics[width=\linewidth]{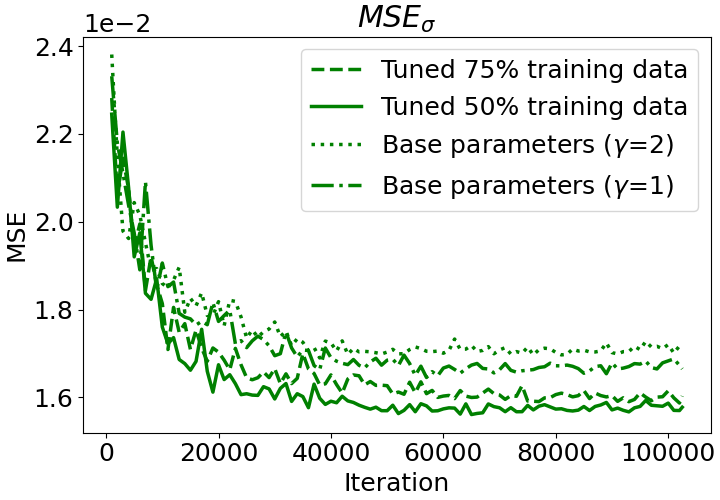}%
        \includegraphics[width=\linewidth]{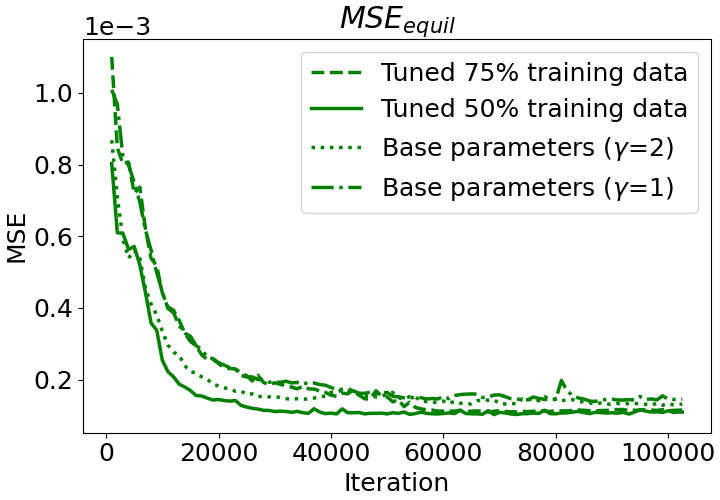}
        \caption{$Tan^{-1}$}
        \label{graph_tuning_TN}
\end{subfigure}

\caption{Tuned hyper-parameters from Table \ref{tab:Data_size_tune_params} and base Pix2Pix hyper-parameters for the (a) baseline, (b) simple addition, (c) sigmoid, and (d) $tan^{-1}$ methods. Results were generated from the validation dataset and used to choose the set of hyper-parameters for final comparison between methods. Note that the y-scale for the sub-figures differ between methods to bring focus on the differences between hyper-parameter sets for a given method (a comparison of all the methods is shown in Figure \ref{fig:final_result}). }
\label{fig:tuning_compare}
\end{figure}

The networks were tuned using 50\% and 75\% of the training data to find the hyper-parameters sets found in Table \ref{tab:Data_size_tune_params}, then trained using all of the training data. Each method was compared to the ``base'' hyper-parameters, or those used in the Pix2Pix paper \cite{isola2018} (shown in Table \ref{tab:Data_size_tune_params}). Training graphs for the full-sized training dataset from each hyper-parameter set are in Figure \ref{fig:tuning_compare}. The graphs show the average performance for each model across ten training sessions and were generated using validation data. The tuned and base parameter models used the same $\gamma$ value if implemented (Note: $tan^{-1}$ resulted in different $\gamma$ values for the different dataset sizes, which is why there are two base hyper-parameter lines). Generally, models optimized using 50\% or 75\% of the training data resulted in a more balanced performance and lower overall errors than the model using the base parameters. The same learning rate and $\beta$ values were found for the simple addition method, indicating that its optimized parameters may be close to the base parameters. Figure \ref{fig:tuning_compare} shows that the fine-tuned simple addition models reduce the divergence error more so than the simple addition method using the base parameters, but this results in larger stress errors than what was preferred. Thus, the base hyper-parameters were kept for the simple addition method.

The hyper-parameters found using 50\% of the training data are not significantly different than those found using 75\% for the sigmoid and baseline models. The baseline and sigmoid methods had slightly better performance with hyper-parameters found using 75\% of the training data than with 50\%. The baseline method sees a similar stress-equilibrium error trade-off as seen in Figure \ref{fig:L2_vs_Div}, though the parameters from 75\% training data seem to find a good balance between the two errors. The $tan^{-1}$ has slightly better performance with the hyper-parameters found using 50\% of the training data.

\subsection{Network Convergence}
\begin{figure}[H]
\centering
\captionsetup{width=.9\linewidth}
\begin{subfigure}{0.49\textwidth}
    \includegraphics[width=\textwidth]{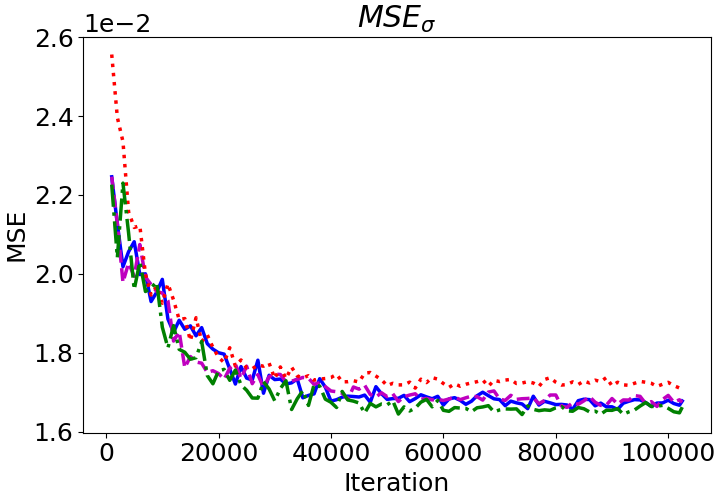}
    \caption{}
    \label{fig:final_mse}
\end{subfigure}
\vspace{0.2in}
\begin{subfigure}{0.49\textwidth}
    \includegraphics[width=\textwidth]{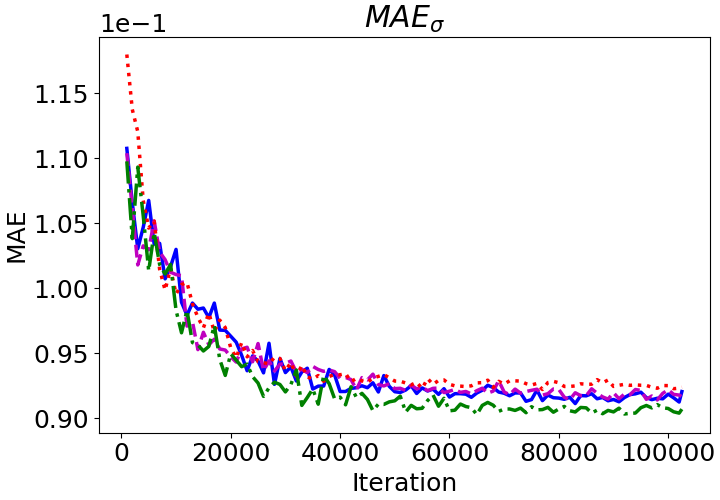}
    \caption{}
    \label{fig:final_mse}
\end{subfigure}
\vspace{0.2in}
\begin{subfigure}{0.49\textwidth}
    \includegraphics[width=\textwidth]{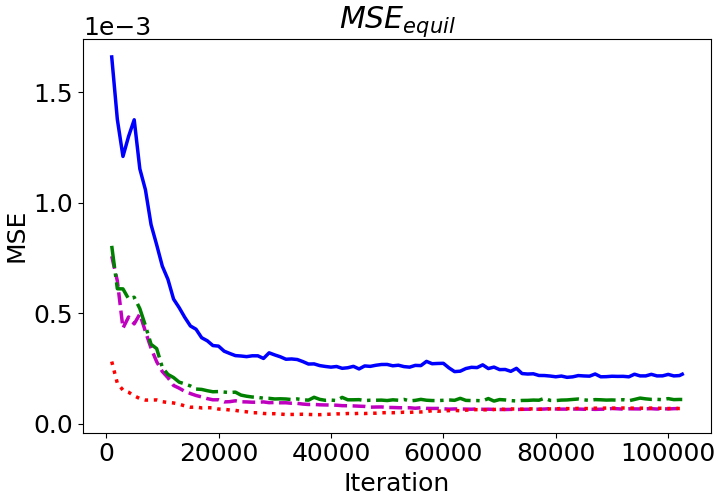}
    \caption{}
    \label{fig:final_div}
\end{subfigure}
\begin{subfigure}{0.49\textwidth}
\hspace{0.5in}
    \includegraphics[width=0.8\textwidth]{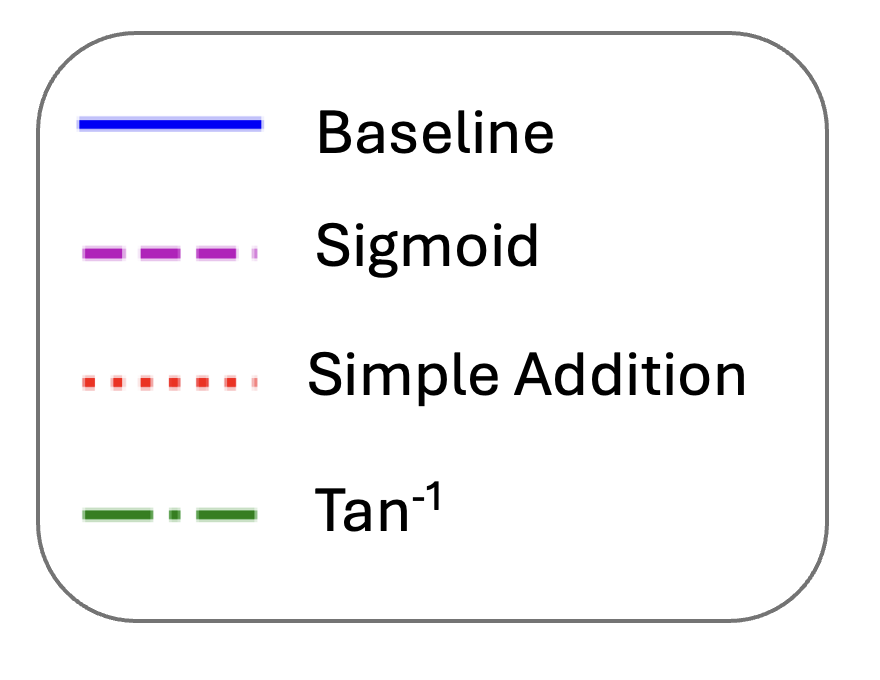}
    \vspace{0.4in}
\end{subfigure}
\caption{The (a) $\text{MSE}_{\sigma}$, (b) $\text{MAE}_{\sigma}$, and (c) $\text{MSE}_{equil}$ for each method during training with the full dataset. Models were saved every 1,000 iterations during training then later tested with validation and test datasets. The graphs are an average across ten training sessions and results were generated with the test dataset from the saved models.}
\label{fig:final_result}
\end{figure}

\begin{figure}
    \centering
    \captionsetup{width=\linewidth}
    \includegraphics[width=\linewidth]{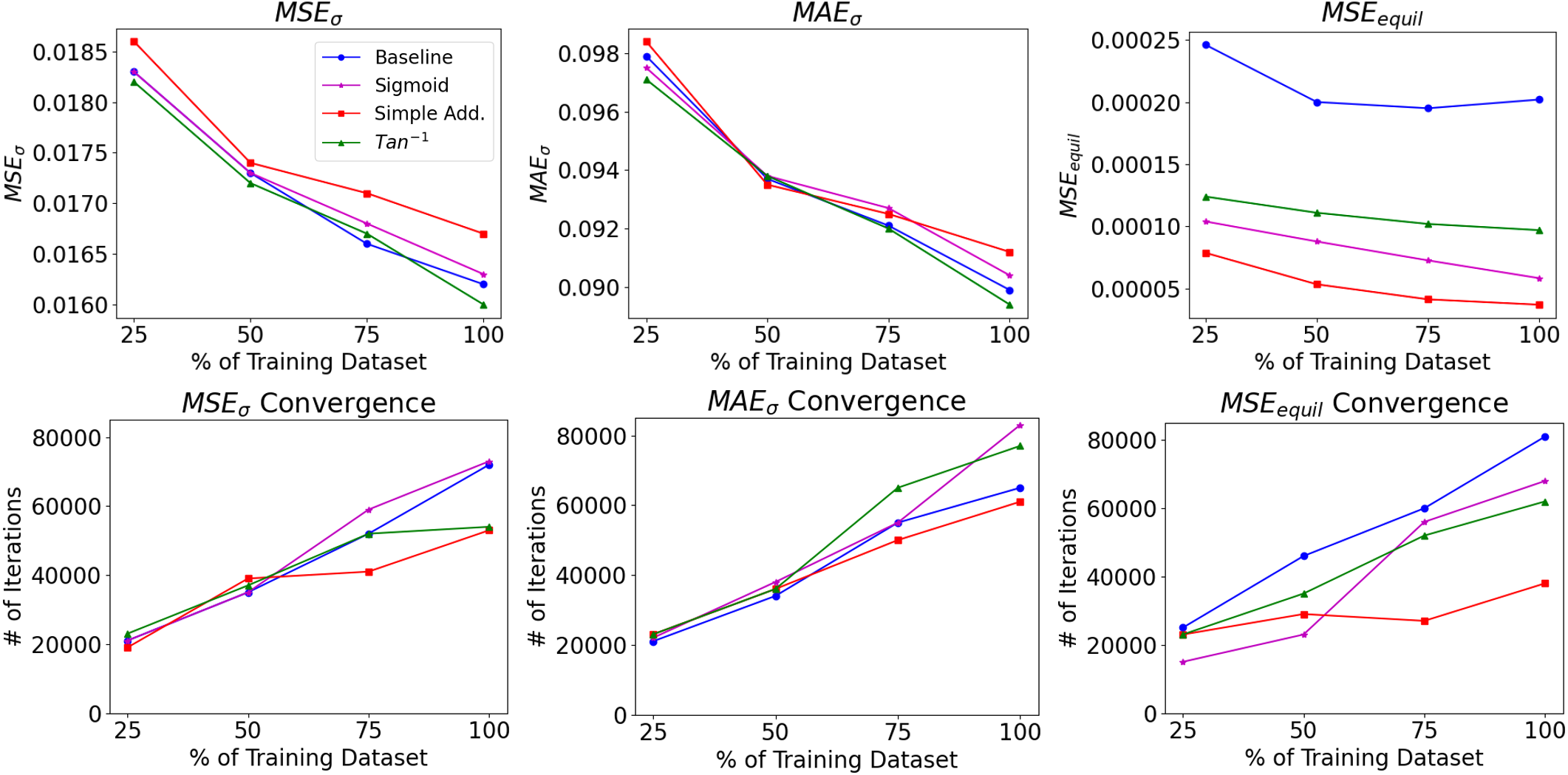}
    \caption{The average lowest error during training for a given metric (first row) and the average iteration a metric will achieve the lowest value (bottom row) with varying dataset sizes. Plots are averaged across the ten training sessions. All plots use the legend shown in the upper left graph.}
    \label{table2_fig}
\end{figure}

The performance throughout training for each method, when trained with the full dataset and using the final hyper-parameters, are plotted together in Figure \ref{fig:final_result}. These plots were generated from the test dataset from the saved checkpoints and averaged across the ten different training sessions. There is a significant improvement in the divergence error throughout training by using a PBR term. The mean absolute error of the stress fields ($\text{MAE}_{\sigma}$) and $\text{MSE}_{\sigma}$ are used to understand low and high-frequency errors, respectively. Figure \ref{table2_fig} shows the average number of iterations to individually reduce each metric and the average lowest error achieved during training for various training dataset sizes (exact values are listed in Table \ref{table:reduce_iter} in the Supplemental Information section). Generally, the errors for all models increase as the dataset size decreases, which is expected. The stress errors across all methods are similar for each dataset size, with the simple addition method performing slightly worse than the rest. 
The divergence error for the baseline does not see a steady decrease in $\text{MSE}_{equil}$ with additional data after 50\%. 
In contrast, all PBR models experience a steady decrease in $\text{MSE}_{equil}$ with more data. In addition, all PBR models reduce $\text{MAE}_{equil}$ in fewer iterations than the baseline model. However, the stress error convergence trends are less clear. The baseline reduced the stress errors before the equilibrium errors (with $\text{MAE}_{\sigma}$ reduced first). The sigmoid and simple addition methods reduce the equilibrium error first then the stress errors (with the $\text{MSE}_{\sigma}$ reduced before $\text{MAE}_{\sigma}$). The $tan^{-1}$ method reduces $\text{MSE}_{\sigma}$, then $\text{MSE}_{equil}$ and lastly $\text{MAE}_{\sigma}$.
To summarize, we do not see any distinct convergence or data efficiency trends when comparing the baseline to the PBR models as a whole regarding stress errors. 
However, there is a clear trend that all PBR methods reduce $\text{MSE}_{equil}$ more than the baseline and converge it more quickly.

\begin{table}[H]
\captionsetup{width=0.9\linewidth}
\caption{Tuned model parameters found using different dataset sizes and the base Pix2Pix parameters for comparison. $\alpha_{D}$ and $\alpha_{G}$ are the learning rates of the $D$ and $G$ network, respectively. 
}
\begin{adjustwidth}{0.4in}{0.1in}
\setlength\tabcolsep{5pt}
\begin{tabular}{l l c c c c c}
\hline\hline
Model & & $\alpha_{D}$ & $\alpha_{G}$ & $D-weight$ & \makecell{L1/L2 weight \\ ($\beta$)} & \makecell{PBR weight \\ ($\gamma$)}\\ [0.5ex] 
\hline
\multirow{ 3}{*}{Baseline}& Base & 0.0002 & 0.0002 & 0.5 & 100(L1) & - \\
& 50\% training & 0.00007 & 0.00025 & 1 & 200(L2) & - \\
 & 75\% training* & 0.0001 & 0.00025 & 0.5 & 200(L2) & - \\
\\
\multirow{ 3}{*}{Sigmoid}& Base & 0.0002 & 0.0002 & 0.5 & 100(L1) & - \\
 & 50\% training & 0.000015 & 0.0002 & 0.5 & 700(L2) & - \\
& 75\% training* & 0.00003 & 0.0002 & 1 & 700(L2) & - \\
\\
\multirow{ 3}{*}{\begin{minipage}[t]{0.1\columnwidth}%
Simple Addition
\end{minipage}}& Base* & 0.0002 & 0.0002 & 0.5 & 100(L1) & 50 \\
& 50\% training & 0.0002 & 0.0002 & 0.7 & 100(L2) & 50 \\
& 75\% training & 0.0002 & 0.0002 & 1 & 100(L2) & 50 \\
\\
\multirow{ 3}{*}{$Tan^{-1}$}& Base & 0.0002 & 0.0002 & 0.5 & 100(L1) & 2 \\
& 50\% training* & 0.0001 & 0.0003 & 0.8 & 300(L2) & 2 \\
 & 75\% training & 0.0002 & 0.0002 & 0.9 & 200(L2) & 1 \\
\hline
\end{tabular}
\begin{tablenotes}
      \small
      \item *chosen set of parameters
    \end{tablenotes}
\label{tab:Data_size_tune_params}
\end{adjustwidth}
\end{table}

\clearpage

\subsection{Network Performance}
Table \ref{table:mse} shows the $\text{MSE}_{\sigma}$, $\text{MAE}_{\sigma}$, and $\text{MSE}_{equil}$ from the model's best-performing iteration, averaged across the 10 training sessions. The best iteration was chosen based on the $\text{MSE}_{\sigma}$ since this seemed to produce the best overall model at little expense to the $\text{MAE}_{\sigma}$ and $\text{MSE}_{equil}$ values. Choosing based on $\text{MSE}_{equil}$ resulted in a model having significant degradation in stress field errors. Models chosen from the best $\text{MAE}_{\sigma}$ performance resulted in similar $\text{MSE}_{\sigma}$ values but increased $\text{MSE}_{equil}$. The stress errors are similar across all methods, with the simple addition method having slightly higher than, and $tan^{-1}$ and sigmoid having about the same error as the baseline method for all dataset sizes. All PBR methods reduce the divergence error compared to the baseline method. 

The errors of the singular best-performing model from the 10 training sessions are shown in Table \ref{table:best} and were trained on the full-sized training dataset. The PBR methods still have the lowest divergence errors, though the baseline method has the lowest stress error. The PBR models seem to more closely resemble the stress error averages shown in Table \ref{table:mse}. This may indicate that the baseline method has a wider range of performance across training sessions, whereas the PBR methods seem to be closer to their average performance.

\begin{table}[H]
\captionsetup{width=\linewidth}
\caption{Metrics when the model is chosen based on the best $\text{MSE}_{\sigma}$ performance, averaged across ten training sessions. The value following the $\pm$ indicates the standard deviation of a metric for predictions across the test dataset  (across all training sessions).}
\centering
\setlength\tabcolsep{10pt}
\begin{tabular}{l l c c c}
\hline\hline
& & $\text{MSE}_{\sigma}$ &  $\text{MAE}_{\sigma}$ & $\text{MSE}_{equil}$ \\ [0.5ex]
\hline
\multirow{ 4}{*}{Baseline} & 25\% &  0.0183 $\pm$0.0384 & 0.0979 $\pm$0.0777 & 3.02e-4 $\pm$1.22e-4\\
& 50\% & 0.0173 $\pm$0.0388 & 0.0939 $\pm$0.0771 & 2.42e-4 $\pm$6.43e-5\\
& 75\% & 0.0166 $\pm$0.0368 & 0.0923 $\pm$0.0759 & 2.42e-4 $\pm$8.77e-5\\
& 100\% & 0.0162 $\pm$0.0384 & 0.0902 $\pm$0.0767 & 2.19e-4 $\pm$6.45e-5\\
\\
\multirow{ 4}{*}{Sigmoid} & 25\% &  0.0183 $\pm$0.0385 & 0.0977 $\pm$0.0776 & 1.26e-4 $\pm$3.50e-5\\
& 50\% & 0.0173 $\pm$0.0392 & 0.0939 $\pm$0.0768 & 1.05e-4 $\pm$2.89e-5\\
& 75\% & 0.0168 $\pm$0.0382 & 0.0928 $\pm$0.0767 & 8.65e-5 $\pm$3.29e-5\\
& 100\% & 0.0163 $\pm$0.0388 & 0.0908 $\pm$0.0768 & 6.62e-5 $\pm$2.30e-5\\
\\
\multirow{ 4}{*}{\begin{minipage}[t]{0.1\columnwidth}
    Simple Addition
\end{minipage}}& 25\% &  0.0187 $\pm$0.0407 & 0.0985 $\pm$0.0788 & 8.65e-5 $\pm$2.48e-5\\
& 50\% & 0.0174 $\pm$0.0412 & 0.0936 $\pm$0.0782 & 6.72e-5 $\pm$2.12e-5\\
& 75\% & 0.0171 $\pm$0.0399 & 0.0928 $\pm$0.0781 & 5.24e-5 $\pm$2.02e-5\\
& 100\% & 0.0167 $\pm$0.0384 & 0.0917 $\pm$0.0772 & 5.36e-5 $\pm$2.24e-5\\
\\
\multirow{ 4}{*}{$Tan^{-1}$} & 25\% &  0.0182 $\pm$0.0381 & 0.0971 $\pm$0.0781 & 1.27e-4 $\pm$3.44e-5\\
& 50\% & 0.0171 $\pm$0.0378 & 0.0940 $\pm$0.0763 & 1.14e-4 $\pm$3.43e-5\\
& 75\% & 0.0167 $\pm$0.0382 & 0.0921 $\pm$0.0767 & 1.12e-4 $\pm$3.35e-5\\
& 100\% & 0.0160 $\pm$0.0375 & 0.0898 $\pm$0.0757 & 1.07e-4 $\pm$2.99e-5\\
\hline
\end{tabular}
\label{table:mse}
\end{table}

\begin{table}[H]
\captionsetup{width=0.9\linewidth}
\caption{The best-performing model of the ten training sessions for each method. The model chosen was based on the lowest $\text{MSE}_{\sigma}$ and was trained on the full-sized dataset. The value following the $\pm$ indicates the standard deviation of a metric (for just the best performing training session).}
\centering
\setlength\tabcolsep{10pt}
\begin{tabular}{l c c c c}
\hline\hline
&  $\text{MSE}_{\sigma}$ &  $\text{MAE}_{\sigma}$ & $\text{MSE}_{equil}$ & Iteration\\ [0.5ex]
\hline
Baseline & 0.0150 $\pm$0.0364 & 0.0863 $\pm$0.0739 & 2.38e-4 $\pm$5.82e-5 & 47000\\
Sigmoid & 0.0157 $\pm$0.0399 & 0.0883 $\pm$0.0762 & 5.62e-5 $\pm$2.01e-5 & 97000\\
Simple Addition & 0.0160 $\pm$0.0373 & 0.0890 $\pm$0.0756 & 3.94e-5 $\pm$1.21e-5 & 34000\\
$Tan^{-1}$ & 0.0157 $\pm$0.0376 & 0.0888 $\pm$0.0748 & 9.96e-5 $\pm$2.81e-5 & 53000\\
\hline
\end{tabular}
\label{table:best}
\end{table}

\begin{figure}[H]
\centering
\captionsetup{width=.9\linewidth}
\includegraphics[width=0.9\textwidth]{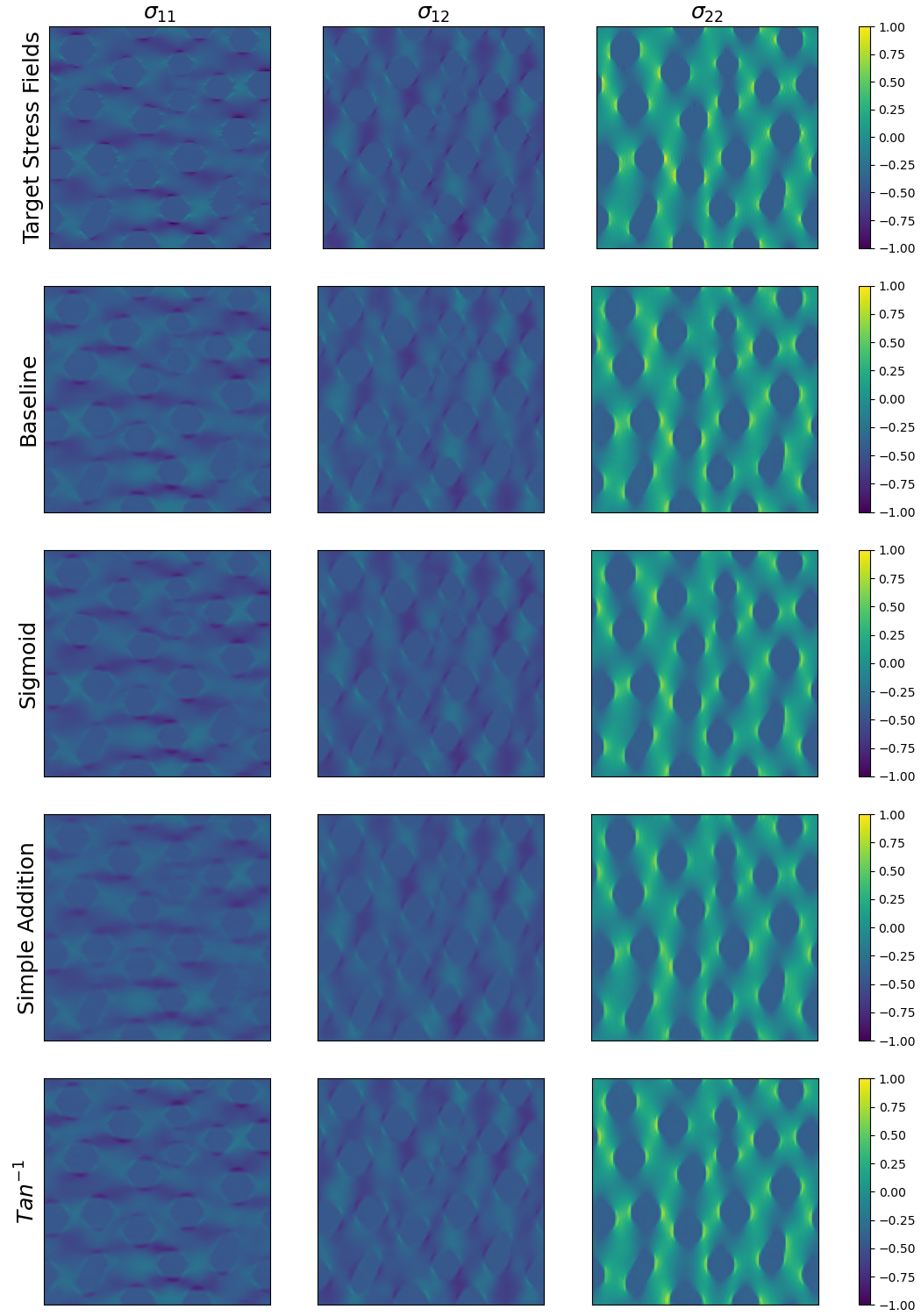}
\caption{Generated stress fields from the best performing model of the ten training sessions for each method. }
\label{stress-fields all}
\end{figure}

\begin{figure}[H]
\centering
\captionsetup{width=.9\linewidth}
\includegraphics[width=1\textwidth]{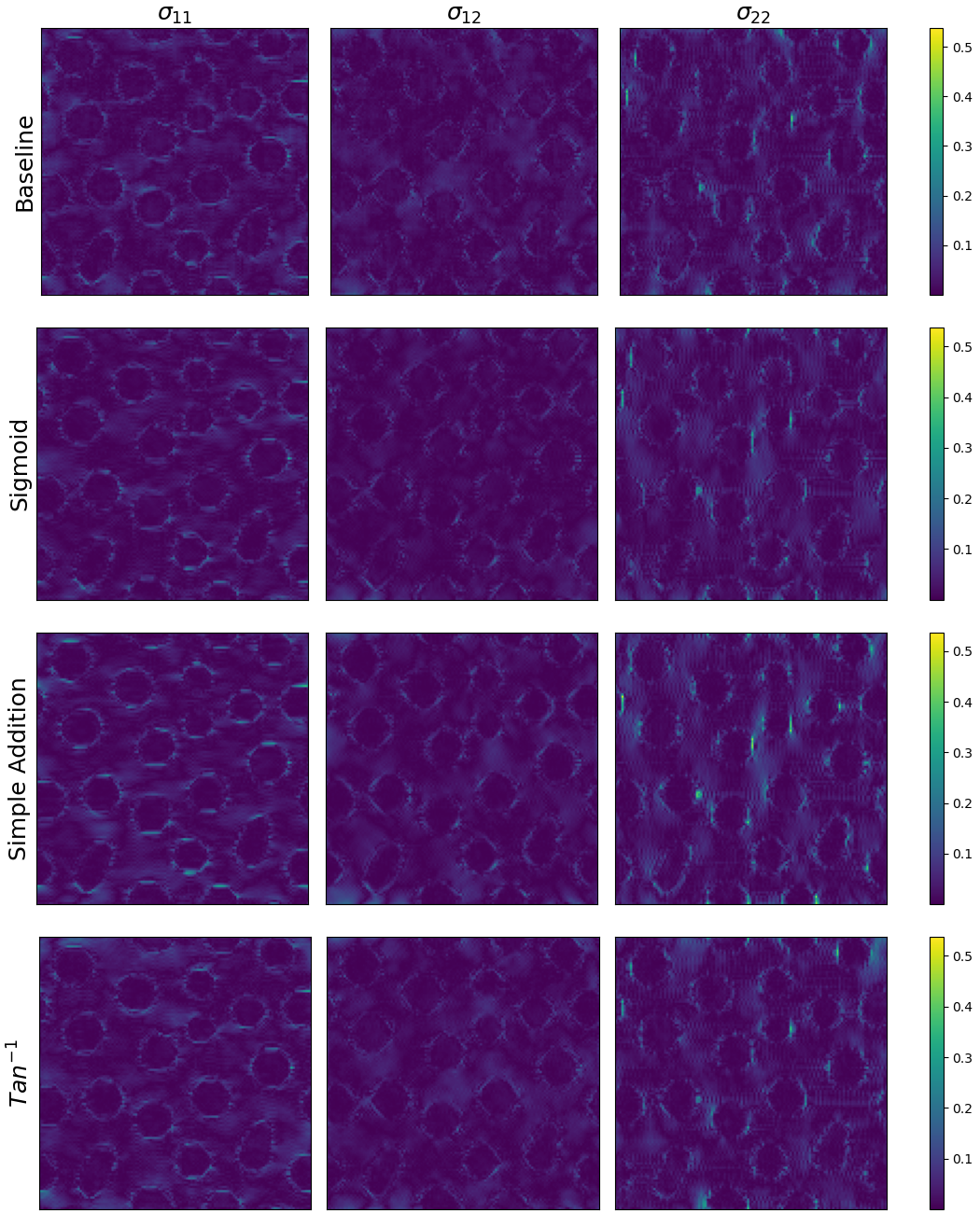}
\caption{The absolute error of the stress fields from Figure \ref{stress-fields all}.}
\label{stress field error}
\end{figure}

Stress fields generated from the models listed in Table \ref{table:best} are shown in Figure \ref{stress-fields all}. The corresponding absolute differences are shown in Figure \ref{stress field error}. The predicted stress fields for each method have high similarity to each other and to the target stress fields. The difference plots are also similar across the different methods, with the 11- and 22-directions having greater errors at the phase interface than the 12-direction. The models do not capture the Gibbs oscillations that are most noticeable in the 22-direction of the target stress fields, and their absence is also noticeable in the difference plots.

\begin{figure}[H]
\centering
\captionsetup{width=\linewidth}
\includegraphics[width=0.53\textwidth]{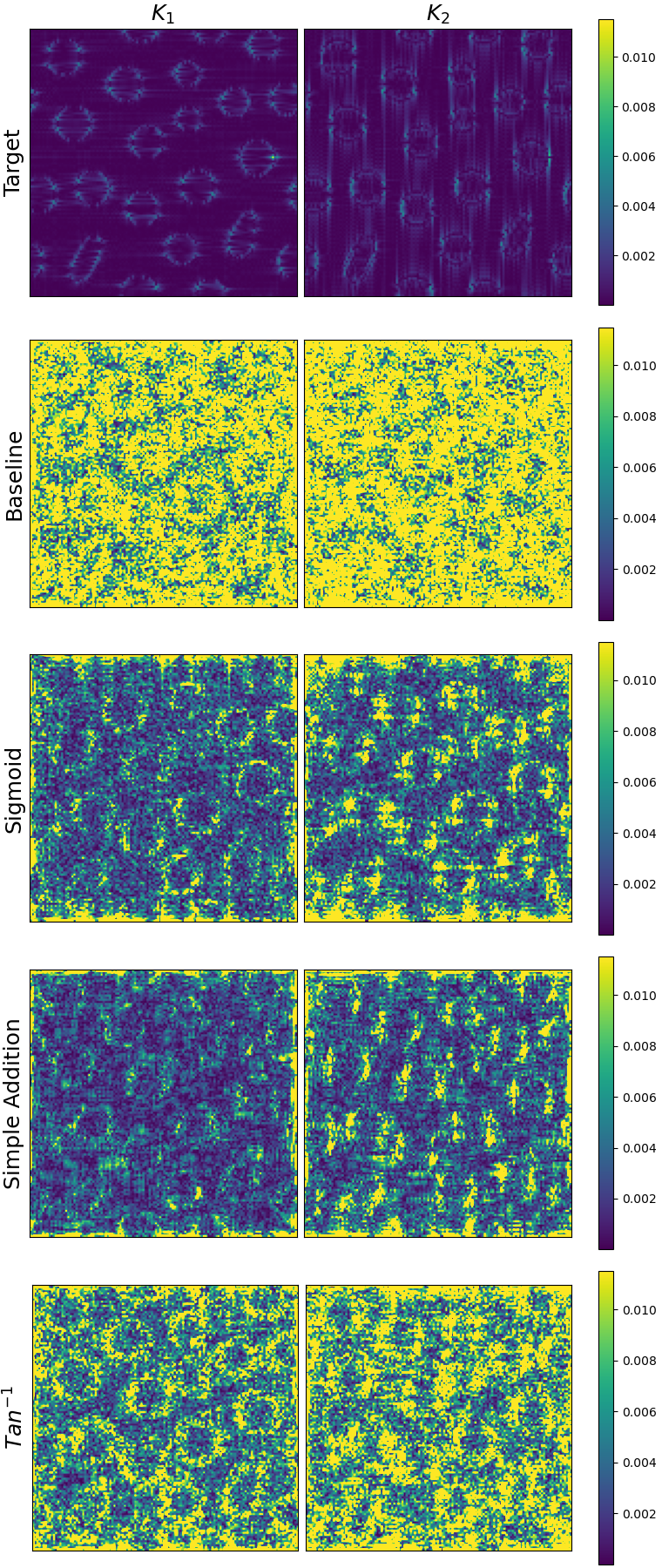}
\caption{The absolute value of the divergence fields calculated from the stress shown in Figure \ref{stress-fields all}. The divergence fields are scaled to the target divergence fields' minimum and maximum values. Yellow pixels indicate a value greater than or equal to the target's largest value.}
\label{div field error}
\end{figure}

The absolute value of the divergence fields calculated from generated or target stress fields are shown in Figure \ref{div field error} (calculated from the same stress field in Figure \ref{stress-fields all}). The divergence fields are scaled to the minimum and maximum value of the absolute value of the target divergence fields. For the generated divergence fields, anything that is a yellow pixel indicates a value that is greater than the largest value of the target stress field (i.e., the target divergence field's largest error). The baseline method's divergence is mostly yellow pixels, showing that this method is not converging anywhere close to the target's equilibrium solution. The PBR methods have a much smaller number of pixels outside of the target divergence range and further demonstrates the effectiveness of the PBR methods in enforcing the equilibrium criteria. The sigmoid and simple addition methods are closest to the equilibrium convergence criteria from the target dataset. The $tan^{-1}$ method has more values outside of the target range but reduces the number in comparison to the baseline method. There are greater errors at the phase boundary for target and generated stress fields. For the target divergence fields, the greater deviation from equilibrium at the phase boundary is a result of the Gibbs oscillations from the high elastic contrast of the phases. For the generated divergence fields, there are greater errors at the phase interface in the stress fields (Figure \ref{stress field error}), which likely contributes to the greater errors at the interfaces in the divergence fields.

\section{Discussion}
Taken as a whole the above results highlight the need to carefully consider how incorporating a physics based loss may affect training and network performance. However, there are several key points that warrant further discussion since they could provide insight on best practices for application of DL in science and engineering more generally.

The gradients of a loss function or regularization term must also be considered when formulating the loss. The network updates the weights of the model by calculating the derivative of the loss for every weight, indicating the direction that weight should be updated to minimize the loss. Very large or small gradients could lead to the known problem of exploding or vanishing gradients, respectively. This leaves the network incapable of updating its weights. The sigmoid function is a commonly used activation function for its stable gradients. The simple addition method method has gradients similar to $|x|$ and results in stable training. One method that was initially tested but ultimately excluded in this study was unstable during training for this reason. In this rejected method an additional regularization term was added to the generator loss comparing the target RMS of divergence to the predicted RMS of divergence: $ln(\frac{RMS_{i,fake}}{RMS_{i,target}})$. For this term, as the RMS of divergence from predicted stress fields approached values close to the RMS of divergence from target stress fields, the gradients will become larger due to the nature of the natural log function, leading to exploding gradients. The $tan^{-1}$ method was implemented instead due to its more reasonable gradients by comparison.

With the exception of the simple addition method, each model required a change in learning rates and loss weights with respect to the original Pix2Pix parameters to achieve a better-performing network. Furthermore, each method required a unique set of hyper-parameters relative to the differing regularizers. This is a key result from this paper, strongly suggesting that when incorporating PBR, careful tuning of the hyper-parameters is required for each regularization method considered. This shows that learning 

\begin{figure}[H]
\centering
\captionsetup{width=.9\linewidth}
\begin{subfigure}{0.49\textwidth}
    \includegraphics[width=\textwidth]{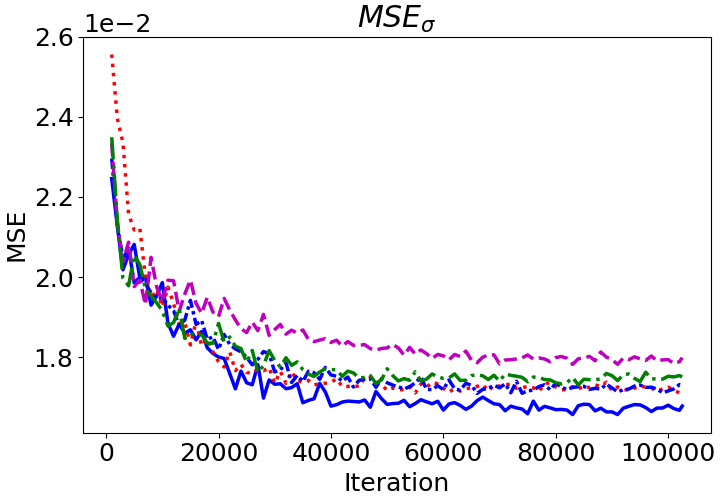}
    \caption{}
    \label{fig:final_mse}
\end{subfigure}
\vspace{0.2in}
\begin{subfigure}{0.49\textwidth}
    \includegraphics[width=\textwidth]{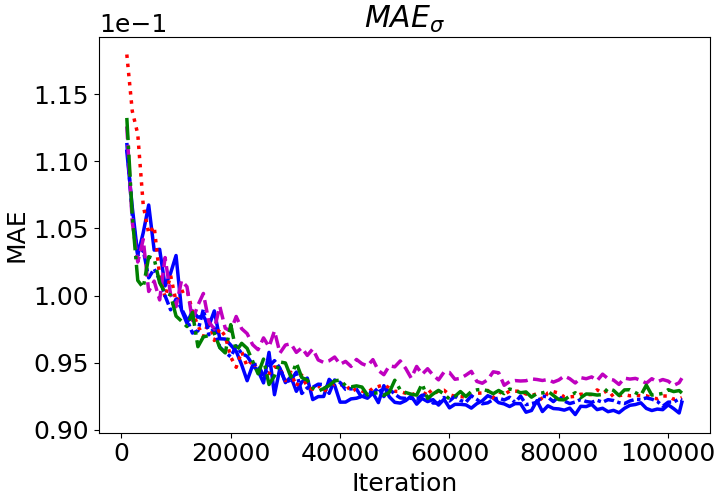}
    \caption{}
    \label{fig:final_mse}
\end{subfigure}
\vspace{0.2in}
\begin{subfigure}{0.49\textwidth}
    \includegraphics[width=\textwidth]{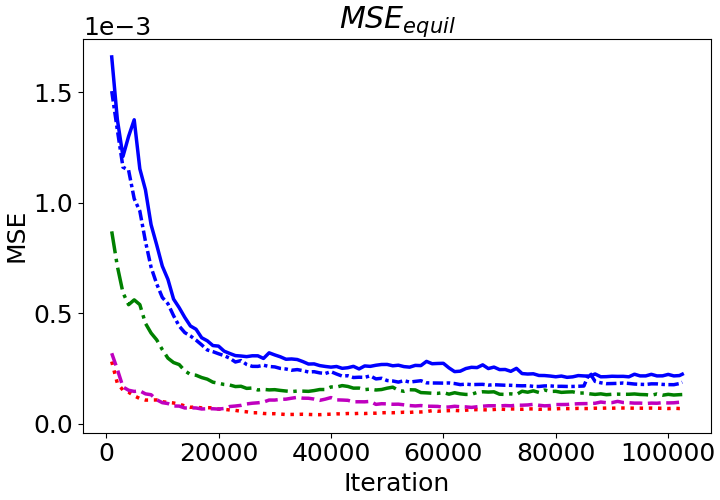}
    \caption{}
    \label{fig:final_div}
\end{subfigure}
\begin{subfigure}{0.49\textwidth}
\hspace{0.5in}
    \includegraphics[width=0.8\textwidth]{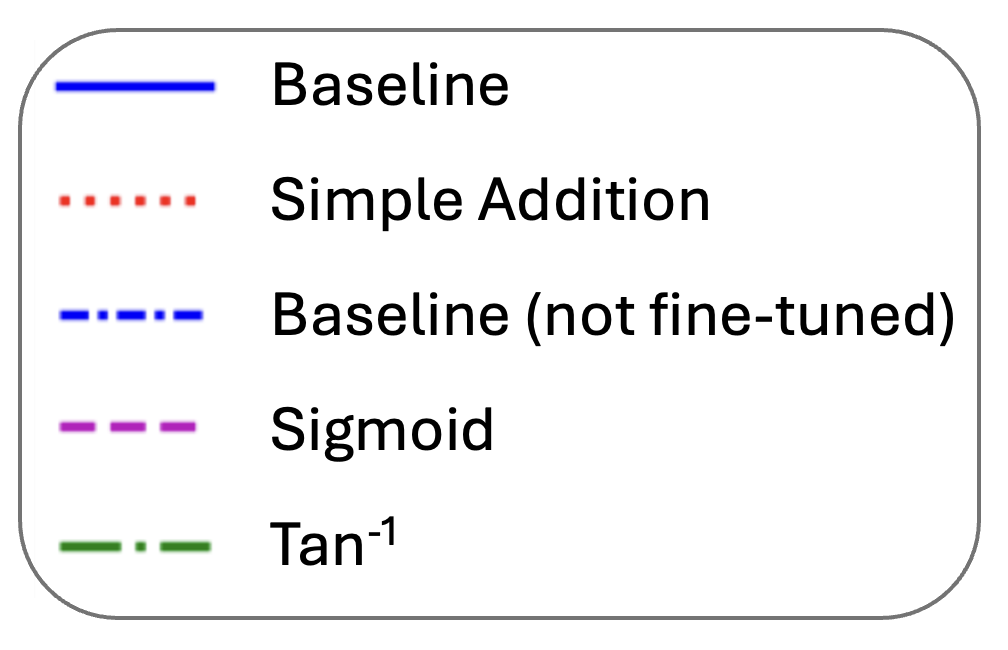}
    \vspace{0.4in}
\end{subfigure}
\caption{Each method with the base Pix2Pix hyper-parameters (i.e. only tuning $\gamma$, if applicable) throughout training, with the full dataset. The baseline method has the fine-tuned (solid line) and not-tuned (dash-dotted line) versions for comparison.}
\label{all_not_tuned}
\end{figure}

\noindent rates and loss weights are not ``one-size-fits-all'' for a given learning task, and fine-tuning needs to be done when the loss function is changed.

This result builds on those found by Li et al. (2020) \cite{li2020rethinking}. When the dataset is altered, the optimization problem changes and consequently the loss landscape is also altered. As Li et al. (2020) demonstrates, this requires different effective learning rates in the application of transfer learning with differing dataset domains. Materials scientists commonly use networks trained on natural images and transfer the network to materials science-based applications \cite{CHOWDHURY2016,li2018transfer,Lubbers2017,Cohn_2021,goetz2022addressing}, but they should take care to carefully tune and re-adjust regularization and learning rate hyper-parameters \cite{li2020rethinking,goetz2022addressing}. While transfer learning was not used in this work, our results lead to an implied message about transfer learning. Learning rate and regularization weight adjustments were needed for improved performance in the baseline model that had a different dataset than those used in the original Pix2Pix model. This re-enforces the notion that different data domains having different optimization tasks require different hyper-parameters. In addition, altering the loss function through PBR changes the learning task and will also require changes in learning rates or loss weights for optimal network performance, even with the same dataset. Transfer learning may still offer benefits, particularly in cases where training data might be limited; however, practitioners should be cautioned that the resulting model may not be optimal. Further users should carefully balance the benefits of starting with a pre-trained model against the effort required for tuning the learning and loss parameters when introducing new data sources.  

The importance of hyper-parameter optimization is further reinforced by Figure \ref{all_not_tuned}, which shows the $\text{MSE}_{\sigma}$, $\text{MAE}_{\sigma}$, and $\text{MSE}_{equil}$ during training from the networks using the base Pix2Pix hyper-parameters, as well as the fine-tuned baseline network for comparison. By only fine-tuning $\gamma$ (if implemented), the PBR losses reduced $\text{MSE}_{equil}$ regardless, but the $\text{MSE}_{\sigma}$ and $\text{MAE}_{\sigma}$ metrics will suffer without fine-tuning and have larger stress errors than the fine-tuned and not-tuned baselines. This renders the PBR term ineffective in comparison to the baseline. From Figure \ref{fig:final_result}, tuning the PBR models is likely to reduce the stress errors or maintain similar stress errors to the baseline while still reducing the equilibrium error. Thus, the accurate documentation and reporting of hyper-parameters in publications and presentation is just as important as reporting network architecture in creating reproducible models and is equivalent to reporting calibrations or other standards used in experiments. More importantly, any changes in hyper-parameters necessary to reproduce published results should be documented and highlighted in any accompanying code. 

This work shows that learning rate affects the overall performance of a PBR model and is not universal across different methods. This reinforces the argument that learning rates should be reported when proposing a loss for better reproducibility and comparisons across methods. We should also clarify that the purpose of this work was not to identify a procedure or process for optimizing hyper-parameters for physics regularized deep learning models. The networks could potentially be further optimized by using methods like grid or random search for hyper-parameter optimization \cite{liashchynskyi2019grid}. A process for determining optimal parameters is still largely an open question for the machine learning community.

A loss competition was observed during the tuning process between the stress field and equilibrium losses. As previously noted, this competition between several losses has been seen in other studies. However, similar trends between the stress and equilibrium errors were even observed in the baseline model. In general, as the baseline $\text{MSE}_{\sigma}$ reduces, the $\text{MSE}_{equil}$ increases and vice versa (see Figures \ref{fig:tuning_compare} and \ref{all_not_tuned}). This indicates that the competition may not only be a result of multiple loss terms, but from optimizing a network to reduce multiple metrics in general. This also shows that a network can be fine-tuned to favor one metric over another. 

This trade-off between favored metrics has implications for the interpretibility of DL. A common criticism of DL networks are their black box, or non-interpretable nature. This is demonstrated by the baseline network. If the baseline network were able to perform 100\% on predicting the stress field, it would also be, by necessity,  highly successful at enforcing stress equilibrium. In our study, the baseline network reduces the stress error to about the same amount as the PBR models while allowing more deviation from equilibrium than the PBR methods. This suggests that the baseline model may have found a ``loophole'' for reducing the stress error by explicitly considering other features in the data without directly ``learning'', or encoding details of stress equilibrium. This highlights the challenges in interpreting or reverse engineering DL models to extract physics, since the model is in effect doing an ``end run around'' the physical principles. This is possible as the degrees of freedom available to the DL model is orders of magnitude greater than necessary to encode stress-equilibrium principles.  This allows the baseline model to reduce the stress error to an impressive degree without explicit consideration for the equilibrium error, even though the two errors directly correlate.
Adding PBR does not completely eliminate interpretability issues, but does focus a network's attention on user-defined metrics or features that may reduce the possibilities of the network finding a loophole during training. 

The baseline and sigmoid models had the best performance when hyper-parameter optimization was performed using more data. This is likely because with less data the chosen hyper-parameters overfit to that smaller set of data, resulting in worse performance for validation and test datasets. Although, the $tan^{-1}$ model found better parameters when using less training data, without overfitting.
As the training dataset size decreases, Figure \ref{table2_fig} and Table \ref{table:mse} shows the equilibrium error for all PBR methods is still lower than the baseline method. However, the PBR models did not have significant changes in $\text{MSE}_{\sigma}$ or $\text{MAE}_{\sigma}$ compared to the baseline when decreasing dataset size. \textit{This means that these PBR methods cannot necessarily be trained using less data than a model without PBR}. The simple addition and $tan^{-1}$ methods reduced $\text{MSE}_{\sigma}$ more quickly than the baseline model, and the sigmoid method took about the same number of iterations as the baseline to reduce $\text{MSE}_{\sigma}$. All PBR methods reduced $\text{MSE}_{equil}$ more quickly than the baseline model and Figure \ref{fig:final_result} shows that $\text{MSE}_{equil}$ tends to stay near the same value once converged (with the simple addition method increasing slightly at around 60,000 iterations). The sigmoid and simple addition methods reduce the $\text{MSE}_{equil}$ before $\text{MSE}_{\sigma}$, meaning that once $\text{MSE}_{\sigma}$ is reduced, the $\text{MSE}_{equil}$ will be reduced as well, as opposed to the baseline model where $\text{MSE}_{equil}$ has not completely reduced once $\text{MSE}_{\sigma}$ has reduced. Like the baseline, the $tan^{-1}$ reduces $\text{MSE}_{\sigma}$ before $\text{MSE}_{equil}$, but the $tan^{-1}$ method reduces $\text{MSE}_{\sigma}$ before the baseline model. More concisely, the simple addition and $tan^{-1}$ methods converged in fewer iterations, and the sigmoid method about the same number of iterations compared to the baseline (using $\text{MSE}_{\sigma}$ as the deciding metric), with the simple addition and sigmoid methods additionally having the equilibrium error already reduced.


The results of this study suggest that  PBR methods do not drastically reduce the number of iterations to convergence, nor do they reduce the amount of data needed to train a network. This result is somewhat counter-intuitive. The need for less training data and faster convergence are commonly stated as points in favor for using physically informed machine learning models. However, upon review by the authors evidence of such claims compared to a conventional network are lacking in literature \cite{Karniadakis2021,sharma2023,hao2023physicsinformed}, with few discussing data efficiency in any depth \cite{wangPIDeepONet,daw2020physics}. More studies focused on data efficiency of physics-informed models compared to conventional ML models needs to be done to better understand the data requirements. Furthermore, without reporting optimally tuned hyper-parameters, such as learning rate and loss weights, the convergence and data needs of physics-informed models are obscured and may be contributing to the conflicting conclusions in literature. In order to make a stronger statement a similar comparison should be conducted with much lager training datasets. It is a possibility that the training dataset, utilized here, was too small to observe differences in performance or hyper-parameter optimization for different dataset sizes. Using a dataset that has tens of thousands of pairs for training would likely be a better comparison, although time-consuming and often unrealistic for materials science investigations. 

The stress errors were similar for all models, with the simple addition method having the largest errors. This could be a result of this PBR method penalizing the network for not having exactly zero divergence fields when the training dataset does not have exactly zero divergence, causing conflicts during network training. It may also indicate that the simple addition method may need more extensive tuning and loss balancing. In contrast, the sigmoid and $tan^{-1}$ loss terms compare the divergence fields to those from the simulated data used for training. These methods essentially tell the network to converge to a similar solution to the FFT solver, encouraging similar errors in the predictions. This results in the sigmoid and $tan^{-1}$ methods having similar stress errors as the baseline model but now better enforce the physical constraints too. However, this does bring up the question of ``Whether the network should replicate errors in the dataset or learn to filter them out during training, regularization, or other feature selection?'' All datasets, simulated or experimental, have some level of noise or bias that the network may learn if it is not removed through some sort of pre-processing step if possible. The user must decide if the network should faithfully reproduce what is represented in the training dataset or have the network ``correct'' bias/artifacts in the dataset through tuning, regularization, feature selection, etc. In this work, the sigmoid and $tan^{-1}$ methods encourage the network to replicate the divergence criteria from the FFT solver, while the simple addition method enforces a different divergence criteria which may also contribute to the greater stress errors this method has since it is deviating from what is represented in the training dataset. 

After training each method ten times on the full dataset, the $tan^{-1}$ is the most likely to result in the lowest stress error, and the simple addition method in the lowest equilibrium error. However, the simple addition method reduces the equilibrium error at the expense of the stress errors. 
The sigmoid's performance lies somewhere in between the simple addition and $tan^{-1}$ methods' performance. The sigmoid method has about the same average stress error as the baseline method and reduces the equilibrium error more than $tan^{-1}$, but not as much as the simple addition method. 
Choosing the best overall model would depend on whether the stress error or equilibrium error is more important without completely disregarding either one. The sigmoid and $tan^{-1}$ methods find the best balance between stress and equilibrium errors. The $tan^{-1}$ method is the best overall model when emphasizing a reduced stress error, while the sigmoid method would be the best if the emphasis is on reducing the equilibrium error. The same could be said for tuning a network. As seen in Figures \ref{fig:tuning_compare} \& \ref{all_not_tuned}, the baseline model could be optimized to slightly favor one metric over the other. The baseline was fine-tuned to reduce stress errors but did so at some expense of equilibrium error. If equilibrium error was to be emphasized, the base parameters for the baseline model could have been kept, although at the expense of stress accuracy. 

Even though $tan^{-1}$ had the smallest stress errors on average across the different training sessions, the baseline method produced the model with the lowest stress errors (with an increase in divergence error). The stress errors from best PBR models are closer to the stress error averages shown in Table \ref{table:mse}, indicating that the baseline method may have a wider range of performance across different training sessions. The variation in performance across different training sessions of identical models is not often discussed but gives important insight into the repeatability of a DL model. The degree of variability between training sessions, is typically not reported in the literature. To the author's best knowledge, comparison of training variability in physics-informed networks has not been previously discussed.  While this study hints at the necessity to evaluate the performance variability of  deep learning networks, particularly physics-informed networks, a more completely study is necessary. However even these preliminary results suggest that practitioners should be skeptical of results that show significant advantages for physics-informed networks when the variability between training runs has not been adequately characterized. Without understanding the variability of deep learning models, it is impossible to address the repeatability of results for future datasets or to begin to consider critical areas for engineering design (e.g. uncertainty quantification) or even speculate how ML models can be utilized for certification and qualification within the context of Integrated Computational Materials Engineering (ICME). 

This work focused on altering the loss function to incorporate physics into the network. Of course, there are other ways to incorporate known physics into a network, such as feature selection (either isolating features or bringing additional information to the data as a featurization step) as input into the network. Ref. \cite{HERRIOTT2020} showed that different input features (grain ID, grain orientation, load vector, Taylor factor, or Schmid factor) result in different accuracies in yield stress predictions from a CNN, demonstrating that careful feature selection of the input affects the accuracy of a network. 
When the physics of a phenomena is not completely understood or complex to encode as a simple partial differential equation or algebraic model, data augmentation is the natural choice. On the other hand, if the physics are well understood and straightforward to model but the training data contains noise or other artifacts, PBR is a more logical approach. However, for most learning tasks the answer is probably somewhere in the middle. Understanding this balance between regularization and featurization of ML models is currently a wide open area ripe for exploration and exploitation by the community.

\section{Conclusions}
Several physics-based regularization (PBR) loss functions enforcing stress equilibrium were implemented on a Pix2Pix network for predicting the stress fields of a high elastic contrast two-phase composite. Learning rates and loss weights were separately optimized for each implementation. Each PBR method and a baseline method required different sets of learning rates and loss weights from each other, showing that when the loss function is altered some fine-tuning is needed to achieve the best performance. Every PBR method significantly reduced the equilibrium error compared to a model without PBR. However, this error reduction is sometimes at the expense of the stress field error. We found no indication that our PBR models could train with less data compared to a model without PBR, but PBR models are likely to converge more quickly due to faster equilibrium error convergence.

\printbibliography 
\section{Supplemental Information}

\begin{table}[H]
\captionsetup{width=.9\linewidth}
\caption{Average number of iterations to reduced metrics with varying training dataset size. Values are averages from the ten training sessions.}
\centering
\setlength\tabcolsep{9.8pt}
\begin{tabular}{l c c c c}
\hline\hline
& \makecell{\% of the training \\ dataset used}& \makecell{$\text{MSE}_{\sigma}$\\ score $\mid$ iteration} &  \makecell{$\text{MAE}_{\sigma}$\\ score $\mid$ iteration} & \makecell{$\text{MSE}_{equil}$\\ score $\mid$ iteration} \\ [0.5ex]
\hline
\multirow{ 4}{*}{Baseline} & 25\% & \makecell{0.0183 $\mid$ 21000}& \makecell{0.0979 $\mid$ 21000} &\makecell{2.46e-4 $\mid$ 25000} \\
& 50\% & \makecell{0.0173 $\mid$ 35000}& \makecell{0.0937 $\mid$ 34000}& \makecell{2.00e-4 $\mid$ 46000}\\
& 75\% & \makecell{0.0166 $\mid$ 52000}& \makecell{0.0921 $\mid$ 55000}& \makecell{1.95e-4 $\mid$ 60000}\\
& 100\% & \makecell{0.0162 $\mid$ 72000}& \makecell{0.0899 $\mid$ 65000}& \makecell{2.02e-4 $\mid$ 81000}\\
\\
\multirow{ 4}{*}{Sigmoid} & 25\% & \makecell{0.0183 $\mid$ 21000}& \makecell{0.0975 $\mid$ 22000}& \makecell{1.07e-4 $\mid$ 15000}\\
& 50\% & \makecell{0.0173 $\mid$ 35000}& \makecell{0.0938 $\mid$ 38000}& \makecell{8.79e-5 $\mid$ 23000}\\
& 75\% & \makecell{0.0168 $\mid$ 59000}& \makecell{0.0927 $\mid$ 55000}& \makecell{7.27e-5 $\mid$ 56000}\\
& 100\% & \makecell{0.0163 $\mid$ 73000}& \makecell{0.0904 $\mid$ 83000}& \makecell{5.84e-5 $\mid$ 68000}\\
\\
\multirow{ 4}{*}{\begin{minipage}[t]{0.1\columnwidth}
    Simple Addition
\end{minipage}} & 25\% & \makecell{0.0186 $\mid$ 19000} & \makecell{0.0984 $\mid$ 23000} & \makecell{7.88e-5 $\mid$ 23000} \\
& 50\% & \makecell{0.0174 $\mid$ 39000} & \makecell{0.0935 $\mid$ 36000} & \makecell{5.35e-5 $\mid$ 29000}\\
& 75\% & \makecell{0.0171 $\mid$ 41000} & \makecell{0.0925 $\mid$ 50000} & \makecell{4.14e-5 $\mid$ 27000}\\
& 100\% & \makecell{0.0167 $\mid$ 53000} & \makecell{0.0912 $\mid$ 61000} & \makecell{3.71e-5 $\mid$ 38000}\\
\\
\multirow{ 4}{*}{$Tan^{-1}$} & 25\% & \makecell{0.0182 $\mid$ 23000}& \makecell{0.0971 $\mid$ 23000}& \makecell{1.24e-4 $\mid$ 23000} \\
& 50\% & \makecell{0.0171 $\mid$ 37000}& \makecell{0.0938 $\mid$ 36000}& \makecell{1.11e-4 $\mid$ 35000}\\
& 75\% & \makecell{0.0167 $\mid$ 52000}& \makecell{0.0920 $\mid$ 65000}& \makecell{1.02e-4 $\mid$ 52000}\\
& 100\% & \makecell{0.0160 $\mid$ 54000}& \makecell{0.0894 $\mid$ 77000}& \makecell{9.70e-5 $\mid$ 62000}\\
\hline
\end{tabular}
\label{table:reduce_iter}
\end{table}

\end{document}